\PassOptionsToPackage{pdfpagelabels=false}{hyperref} 
\documentclass[fleqn,usenatbib]{mnras}
\usepackage{newtxtext,newtxmath}

\usepackage[T1]{fontenc}
\usepackage{ae,aecompl}
\usepackage{graphicx}	
\usepackage{amsmath}	
\usepackage{amssymb}	
\usepackage{subcaption}
\usepackage{caption}
\usepackage[english]{babel}
\usepackage[utf8x]{inputenc}
\usepackage[T1]{fontenc}
\usepackage[font=normal,labelfont=bf]{caption}
\usepackage{amsmath,amsfonts,bm} 
\usepackage{lscape}
\usepackage{booktabs}
\usepackage{xifthen} 
\usepackage[dvipsnames]{xcolor}
\usepackage{graphicx}
\usepackage[normalem]{ulem}
\usepackage{tabularx}
\usepackage{enumitem}

\usepackage{orcidlink}

\title[Blending effects on cosmic shear]{Galaxy blending effects in deep imaging cosmic shear probes of cosmology}

\author[E. Nourbakhsh et al.]
{Erfan Nourbakhsh$^{\orcidlink{0000-0003-3827-4691}}$\thanks{E-mail: \href{mailto:nourbakhsh@ucdavis.edu}{nourbakhsh@ucdavis.edu}}, J. Anthony Tyson$^{\orcidlink{0000-0002-9242-8797}}$,
Samuel J.~Schmidt$^{\orcidlink{0000-0002-5091-0470}}$, and
\newauthor 
The LSST Dark Energy Science Collaboration 
\\ 
Department of Physics and Astronomy, University of California, One
Shields Avenue, Davis, CA 95616
}

\date{Accepted: 2022 February 21; Revised: 2022 February 14; Received: 2021 December 13}

\pubyear{2022}

\begin{document}
\label{firstpage}
\pagerange{\pageref{firstpage}--\pageref{lastpage}}
\maketitle

\begin{abstract}
Upcoming deep imaging surveys such as the Vera C. Rubin Observatory Legacy Survey of Space and Time will be confronted with challenges that come with increased depth. One of the leading systematic errors in deep surveys is the blending of objects due to higher surface density in the more crowded images; a considerable fraction of the galaxies which we hope to use for cosmology analyses will overlap each other on the observed sky. In order to investigate these challenges, we emulate blending in a mock catalogue consisting of galaxies at a depth equivalent to 1.3 yr of the full 10-yr Rubin Observatory that includes effects due to weak lensing, ground-based seeing, and the uncertainties due to extraction of catalogues from imaging data. The emulated catalogue indicates that approximately 12 per cent of the observed galaxies are ``unrecognized'' blends that contain two or more objects but are detected as one. Using the positions and shears of half a billion distant galaxies, we compute shear--shear correlation functions after selecting tomographic samples in terms of both spectroscopic and photometric redshift bins. We examine the sensitivity of the cosmological parameter estimation to unrecognized blending employing both jackknife and analytical Gaussian covariance estimators. An $\sim0.025$ decrease in the derived structure growth parameter $S_8 = \sigma_8 (\Omega_{\rm m}/0.3)^{0.5}$ is seen due to unrecognized blending in both tomographies with a slight additional bias for the photo-$z$-based tomography. This bias is greater than the $2\sigma$ statistical error in measuring $S_8$.
\end{abstract}



\section{Introduction}

With the Legacy Survey of Space and Time\footnote{\url{https://www.lsst.org}} (LSST; \citealt{Ive19}) to be undertaken over the next decade by the Vera C. Rubin Observatory, we will have samples of over 10 billion galaxies.  No longer limited by statistical errors, systematic errors become dominant. For studies of the static sky, there is a class of systematic errors that emerge at low surface brightness. With over 800 images of each sky patch, the LSST will reach ultralow surface brightness.
At 30 mag arcsec$^{-2}$ there are non-trivial challenges with instrument signature removal and sky determination.  Most pipeline processing algorithms oversubtract sky at these levels~\citep{ji+}.
Even if a non-biased sky level can be determined, a related systematic emerges at ultralow surface brightness: an increasing fraction of the field is occupied by galaxy faint halos.  

In astronomical imaging surveys, galaxies are seen as 2D surface brightness objects. For galaxies that project very close to each other on the sky, a dip in surface brightness between them suffices to tag them as ``recognized blends'' and software can attempt to {\it de-blend} the pair or group \citep{scarlet}.  Depending on depth, a certain fraction of galaxies are so blended as to be unrecognized as a blend. Observation of ``unrecognized blends'' yields common measurements for their component galaxies, affecting all quantities that are based on photometry, including most importantly for this study, shape, and photometric redshift. The measurement of galaxy shapes can suffer systematic effects since the galaxies are not individually measured~\citep{blend-review, maccrann2022}.  In ultradeep surveys, a higher fraction of galaxies become unrecognized blends~\citep{dawson2016}. 
The observed galaxies in recent deep photometric surveys such as Dark Energy Survey\footnote{\url{https://www.darkenergysurvey.org}} (DES; \citealt{derose2021}) and the Deep Lens Survey\footnote{\url{http://dls.physics.ucdavis.edu}} (DLS; \citealt{jz}) are extracted from crowded images with a considerable fraction of them overlapping each other.
This is also evident in the Subaru's Hyper Suprime-Cam (HSC; \citealt{hsc18}) deep imaging. The HSC Subaru Strategic Program\footnote{\url{https://hsc-release.mtk.nao.ac.jp}} (HSC-SSP; \citealt{Aihara+}) is regarded as a precursor to LSST with a similar seeing.

At fainter magnitudes, the unrecognized blend fraction grows larger, and the effects of blending become more pronounced as the projected surface number density of the galaxies in a field increases. 
This prompted our investigation of the effects of galaxy blending on estimates of cosmology via a ``realistic'' synthetic catalogue, in the sense that a full {\it N}-body simulation, galaxy numbers and magnitudes, and lensing effects are included.
To make this study feasible, such a catalogue must also include the forward-modelled effects of blending through a process we refer to as {\it emulation}. One can selectively turn on each of the components in the emulated catalogue: seeing (point spread function; PSF) full width half maximum (FWHM), photometric redshift (redshift estimated via multiband photometry), lensing magnification, and various angular and magnitude filtering schemes. However, the current work is devoted to the coupled effects of blending and photometric redshift.

To constrain cosmology with weak lensing (WL) measurements, robust observational results with per cent level precision over a wide range of scales and cosmological parameters are needed \citep{srd}. Exploring shear--shear correlations across tomographic redshift bins is a powerful tool for such studies (e.g. \citealt{derose2021,heymans2021}).
Because these multiple correlations over a large volume jointly have much higher dimensionality than the 2D blending, incorporating this extra information is hoped to minimize biases in cosmological parameter estimation.
The full 3D {\it N}-body simulation used in this study enables us to investigate the impact of unrecognized blending on the cosmological parameter estimation in tomographic analyses of large surveys such as LSST.
We include the effect of unrecognized blending on the photometric redshifts and also study the effect of excluding from the sample the ``catastrophic'' recognized blends where the blend is recognized but its surface brightness profile is severely affected by the nearby galaxies beyond a threshold. This threshold set by ``purity'' is described in Section \ref{recbl}.

We have developed a pipeline, \texttt{BlendSim}, to perform the analysis described here. The paper is organized as follows. We first describe the synthetic sky catalogue and our enhancements in Section \ref{bz}. We then introduce our blending emulation process in Section \ref{blending}.
Section \ref{photozest} explains the method we use to estimate our photometric redshifts and Section \ref{cosmicshearprobe} discusses ``cosmic shear'' as the utilized probe of cosmology and presents our measured two-point correlations for the cases of blending, no blending, true redshifts, and photometric redshifts. Our measured covariance is discussed in Section \ref{covariance}. In Section \ref{cosmoparam} we present the resulting bias of the estimates for cosmological parameters due to blending, and in Section \ref{discussion} we discuss the overall results of this study, and its limitations, in the context of the LSST survey. Appendix \ref{ngl} examines the effect of non-Gaussian likelihood in constraining cosmology using jackknife covariances. Appendix \ref{cosmosis_gcov} presents the results obtained from Gaussian covariance matrices as an alternative to jackknife.

\section{Simulation}\label{bz}
We wish to compare cosmological constraints based on cosmic shear for non-blended vs blended galaxies and with true (hereinafter spectroscopic) redshifts vs photometric redshifts in a large simulation. Given that we seek to understand the systematic effects of blending in a survey of billions of galaxies (LSST), we must rely on a well-defined sample of sufficiently large size whose statistical noise is smaller than the sub-per cent systematic errors we wish to resolve. The simulation must be based on a credible model for large-scale structure, have a well-understood mechanism for tagging the resulting galaxies, and span a sufficient redshift range to support the cosmological analysis. Moreover, it is preferred that the simulation has more than one realization that helps with reducing the statistical noise in our analysis. Having more than one realization is also a check on reproducibility. As described in the subsection below, a catalogue of billions of faint galaxies called \texttt{Buzzard} meets our requirements.

A full image-based approach is prohibitively time-consuming for billions of galaxies. 
We bypass this procedure by simulating images for only a subset of the sky where the galaxy number density is high enough to potentially allow blending to take place.  Hence, these galaxies are sufficiently close on the sky to be either recognized or unrecognized blend candidates. The catalogue used to generate these images consists of a table of galaxy properties such as position, size, ellipticity, colours, redshift, and gravitational lens shear produced by a dark matter power spectrum that evolves with redshift.

For our study, it is important to have a synthetic sky catalogue that is photometrically complete for all galaxy types to a limiting magnitude faint enough to enable realistic blending emulation and analysis. That is, the simulation must include a sufficiently high number density of galaxies so that when convolved with a seeing PSF, the surface density of galaxies in a fiducial ``gold sample'' catalogue (where the minimum flux S/N for point sources is 20) results in noticeable blending with galaxies up to $\sim$2 mag fainter. This is expected for the ``LSST gold sample'' according to \cite{dawson2016}.  The 2 mag buffer is a result of the product of the galaxy number--flux relation (for some spectral band) and the corresponding flux. The number-flux relation for galaxies in red passbands is much shallower than euclidean (stars) due to cosmological effects. Including even fainter galaxies does not have a significant impact on the apparent shape or overall flux of the blends.

\subsection{Buzzard mock catalogue}\label{bzmock}
For this study, we use  data from the \texttt{Buzzard} simulation suite \citep{derose}, a full 3D cosmological simulation, which we then augment in several ways to explore the effects of blending on the inferred cosmology from data. The mock catalogues are available in {\sc healpix}\footnote{Hierarchical Equal-Area isoLatitude Pixelisation of a 2-sphere. See \url{http://healpix.sourceforge.net}.} \citep{healpix05} format through the stand-alone {\sc python} module, \texttt{GCRCatalogs}\footnote{\url{http://github.com/LSSTDESC/gcr-catalogs}}, and are described in detail in \cite{derose}, but we briefly summarize the key aspects below.

\texttt{Buzzard} is a set of synthetic galaxy catalogues tuned for the Dark Energy Survey (DES) but applicable to a variety of large-area photometric and spectroscopic surveys. The simulation employs a set of dark matter-only simulations and then utilizes the \texttt{ADDGALS} algorithm \citep{wechsler21} to paint baryonic galaxies on them. 
This simulation is constructed from a set of three dark matter {\it N}-body simulations run using \texttt{L-GADGET2}, a modified version of \texttt{GADGET2}\footnote{\url{http://wwwmpa.mpa-garching.mpg.de/gadget}} with box lengths ranging from 1 to 4 $h^{−1} \rm Gpc$ from which light-cones were constructed. The positions of mass halos are decorated with galaxies whose type, size, brightness, and redshift reflect those in the DES data. Currently, there are two quarter-of-the-sky realizations available in total through \texttt{GCRCatalogs} . 
Weak lensing shears are measured at galaxy positions using an algorithm (\texttt{CalcLens}; \citealt{calclens}) appropriate for full sky arcsecond ray tracing. Galaxy magnitudes and shapes include the impact of shear and magnification of apparent magnitudes in many filter sets including LSST bands. 
Each \texttt{Buzzard} realization has about 3 billion galaxies before any sample cuts spanning an area of 10\,313 square degrees.
A number of validation tests for internal consistency of \texttt{Buzzard\_v2.0} have been carried out \citep{derose2021}. This is the version of \texttt{Buzzard} used in this study.

The \texttt{Buzzard} mock catalogue is populated with galaxies that have magnitudes and shapes designed to approximately match those observed in the DES imaging out to $z = 2.35$ and to a depth of the DES-like magnitude $r \sim 26$.  To avoid potential biases associated with using deconvolved sizes from the actual DES data, the true half-light radii in \texttt{Buzzard} are drawn from empirical observations of galaxies imaged with the {\it Hubble Space Telescope} ({\it HST}) $F125W$ band (J. DeRose, private communication). The spectral energy distributions (SEDs) are drawn from SDSS Data Release 6 (DR6) spectroscopic sample consisting of approximately $5\times10^5$ galaxies. The observed distribution of SEDs at a given luminosity and spatial galaxy density is modelled in SDSS and then used to assign SEDs to simulated galaxies with similar properties.

\begin{figure}
	\centering
	\includegraphics[width=0.48\textwidth]{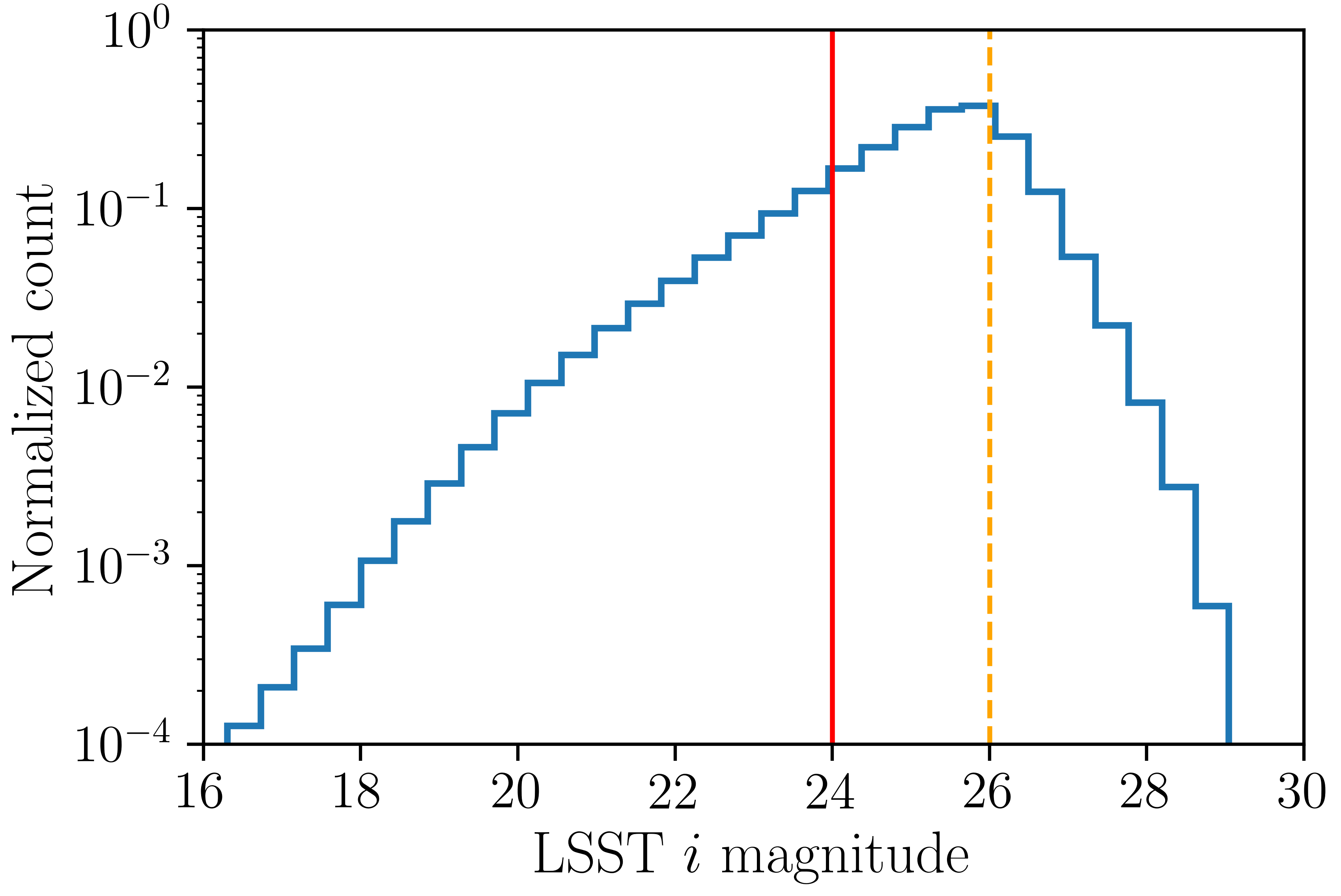}
	\caption{The true $i$ band magnitude distribution of the \texttt{Buzzard} galaxies used in this work. The solid red line indicates our \texttt{Buzzard} cosmology sample cut which ensures $\rm S/N>20$ for point sources brighter than $i=24$. The dashed orange line indicates a cut two magnitudes deeper that we impose in order to remain complete for the faint sample.}
	\label{dndm} 
\end{figure}

\begin{figure}
    \captionsetup[subfigure]{aboveskip=-1pt,belowskip=-1pt}
	\centering
	\begin{subfigure}[a]{0.475\textwidth}
   \subcaption{Spec-$z$}
   \includegraphics[width=1\linewidth]{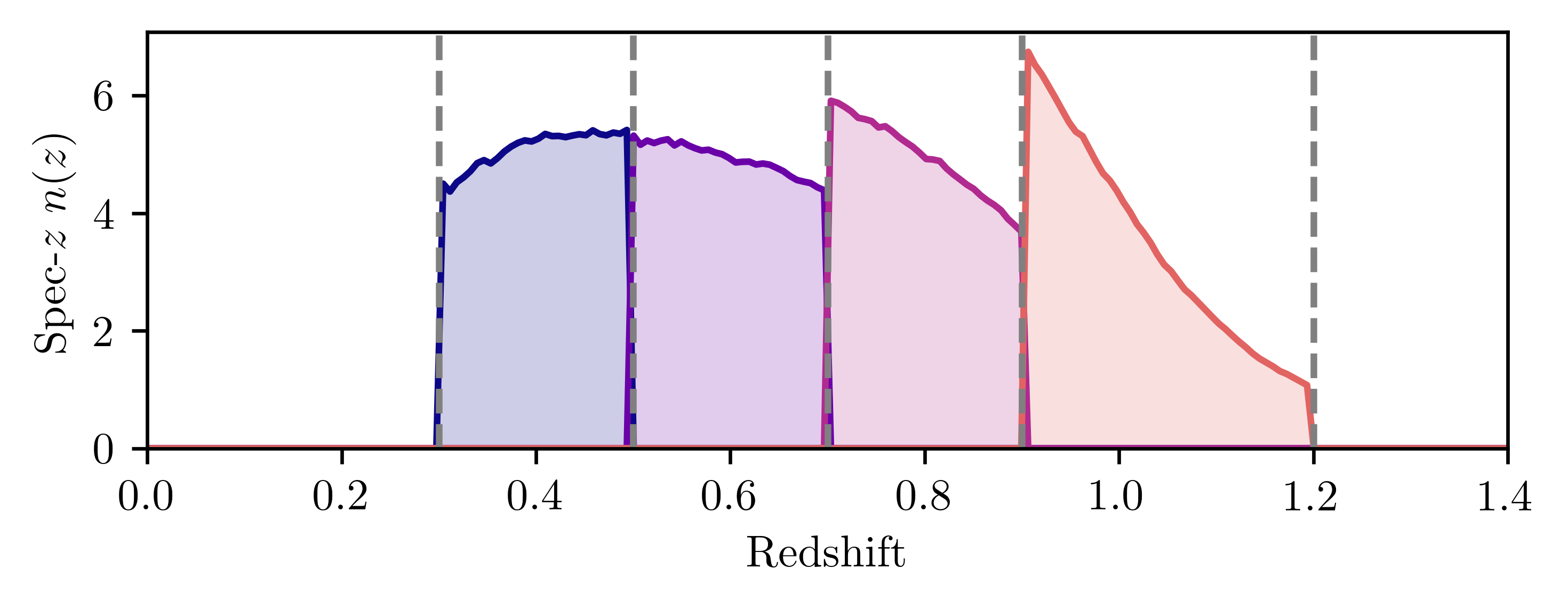}
   \label{dndz_s_specz} 
    \end{subfigure}

	\begin{subfigure}[b]{0.475\textwidth}
   \subcaption{Photo-$z$ point estimates}
   \includegraphics[width=1\linewidth]{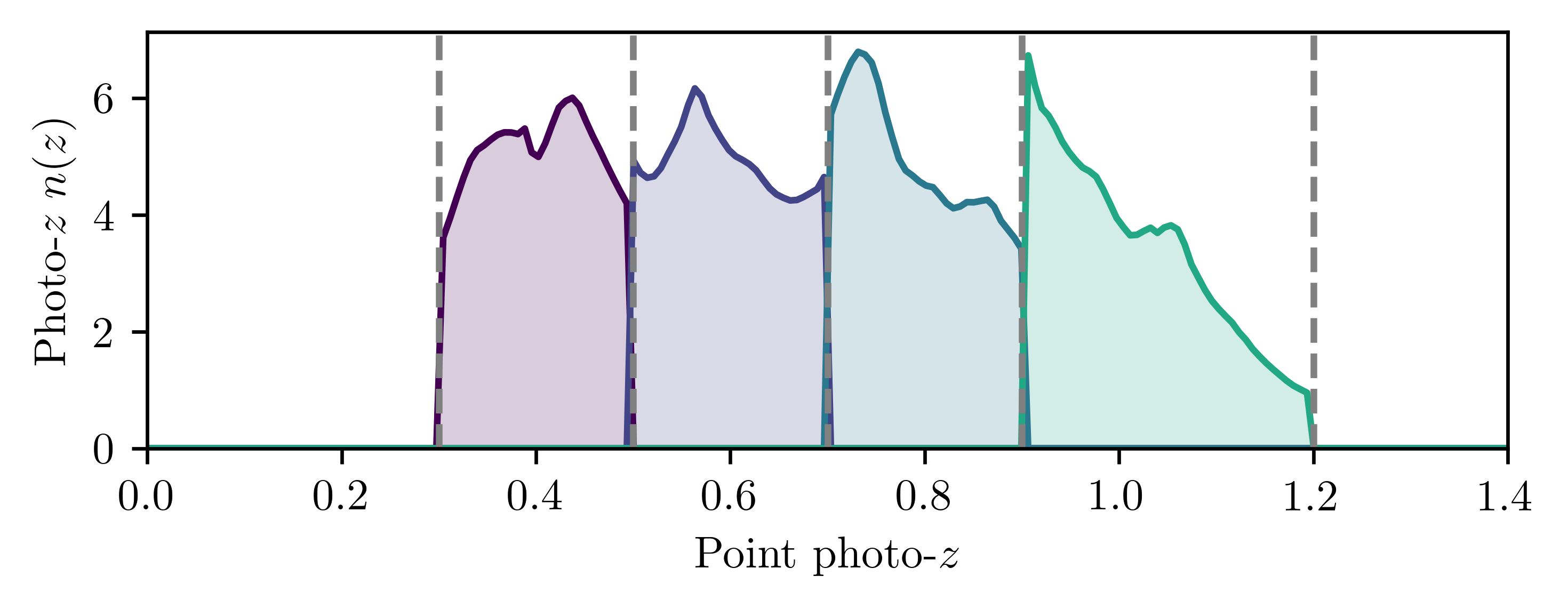}
   \label{dndz_s_photoz} 
    \end{subfigure}

	\begin{subfigure}[c]{0.475\textwidth}
   \subcaption{Photo-$z$ PDF}
   \includegraphics[width=1\linewidth]{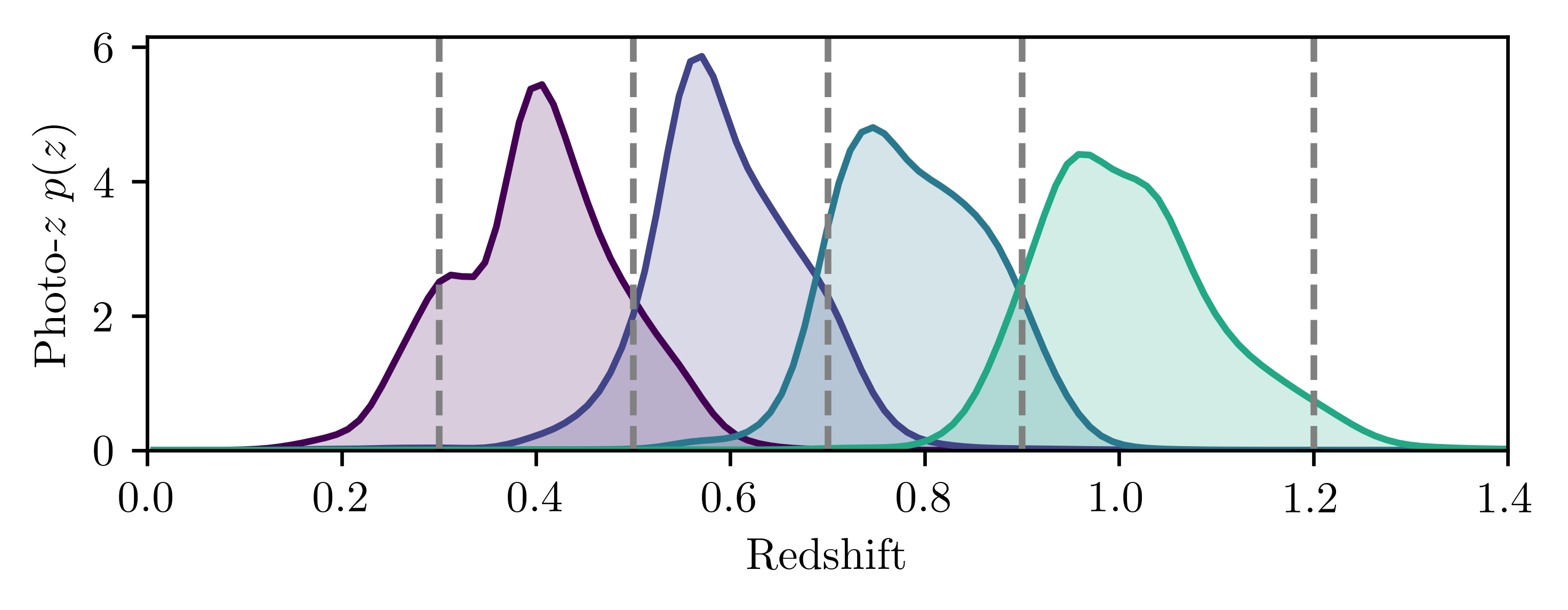}
   \label{dndz_s_photoz_pdf} 
    \end{subfigure}
    
	\caption{The redshift distribution of the non-blended sources with (a) spectroscopic tomography (b) photometric tomography using point estimates (c) photometric tomography using full PDF, $p(z)$ for the \texttt{Buzzard} source sample galaxies used in this work for cosmological inference. The redshift distribution of each tomographic bin is normalized such that its integral over redshift is unity. Panel (c) illustrates that the bins defined via photo-$z$ have ``tails'' due to uncertainties in the redshift estimates. The vertical dashed lines mark the redshift bin edges at $z=0.3,0.5,0.7,0.9$, and $1.2$ used for tomography. These plots are averaged over two realizations of \texttt{Buzzard}. See Table \ref{source_table} for the number counts. We use a different colour palette to show tomographic bins for plot (a) to emphasize that galaxies selected for photo-$z$ tomography (b and c) are selected based on photo-$z$ point estimates and are not the same as those from the spec-$z$ tomographic bins (a) that are selected based on spectroscopic redshifts.}
	\label{tomobins} 
\end{figure}

Fig. \ref{dndm} shows the true distribution of $i$ band magnitude in the mock catalogue. Readers should be aware that the \texttt{Buzzard} sample used in this work is complete to $i \sim 26$, which does not quite reach the depths that we expect to cover in full 10-yr LSST Wide Fast Deep samples.  Thus, we will define our ``gold'' sample with a magnitude cut of $i<24$, and, to avoid confusion with the LSST gold sample where $i<25.3$, we call it \texttt{Buzzard} ``cosmology'' sample from this point on.  In light of the 2 mag buffer justified previously, this upper magnitude cut on the cosmology sample allows us to include galaxies down to $i=26$ for purposes of our blending study while remaining complete.
Further quality cuts described in Section \ref{cosmicshearprobe} are applied to the cosmology sample, which is then referred to as the \texttt{Buzzard} ``source'' sample.  We refer to the sample of all the galaxies with $i<26$ as the \texttt{Buzzard} ``faint'' sample.
Figs \ref{dndz_s_specz}--\ref{dndz_s_photoz_pdf} show the distribution of redshift for the (non-blended) source sample galaxies using the spectroscopic redshift (spec-$z$), photometric redshift (photo-$z$) point estimates and photo-$z$ probability distribution function (PDF), respectively. Section \ref{photozest} provides details regarding the photometric redshift estimation.

Although, the \texttt{Buzzard} simulation has the physics necessary to enable tests of the cosmic shear probe of cosmology, it does not include blending or effects due to the PSF. We discuss how we model these effects in a modified \texttt{Buzzard} catalogue below.  This enables comparison of the resulting blended galaxies catalogue with the original (non-blended) catalogue where the input parameters for all galaxies are known on an individual basis. In other words, even in crowded regions with objects in close proximity we have perfect knowledge of the input fluxes, shapes, and other quantities separately for each galaxy in the non-blended catalogue.  Finally, we can measure the changes encountered as the result of blending in the inferred cosmological parameters.

\subsection{Modified galaxy properties}\label{galprop}

Due to the fact that {\sc healpix} pixels are assigned before lensing was applied, we truncate the sample by removing galaxies within 0.2 degrees of the edge of \texttt{Buzzard}'s footprint\footnote{The region near the outer boundary of the catalogue footprint is affected by both the light-cone and the sky footprint. The lack of galaxies outside the light-cone used to define the simulation sample may bias the modelling of the lensing effects and therefore likely to be discrepant from the rest of the catalogue.} to avoid issues associated with overdense and underdense regions near the edge. This leaves us with $\sim 10\,254$ square degrees of the simulation for this study. We modify the truth-catalogue properties of all faint-sample ($i < 26$) galaxies to more accurately reflect LSST-like samples by convolving the galaxy sizes and shapes with a
\begin{equation*}
\theta_{\rm eff}^{\tilde X} =  \theta_{\rm eff}^{\rm zenith} \tilde X^{0.6} \sim 0.9 ~ \rm arcsec
\end{equation*}
circular Gaussian PSF -- the ``effective'' PSF FWHM value expected at the median airmass of $\tilde X=1.2$ for LSST where $\theta_{\rm eff}^{\rm zenith}=0.8~ \rm arcsec$ is the ``effective'' zenith seeing FWHM for the $i$ band taken from Table 2 of \cite{Ive19}. The reason we use this ``effective'' value is to account for the deviation of the LSST fiducial atmospheric PSF (a von K\'arm\'an type PSF) from the Gaussian shape \citep{angeli16} assumed here.

\begin{figure*}
        \includegraphics[width=1\linewidth]{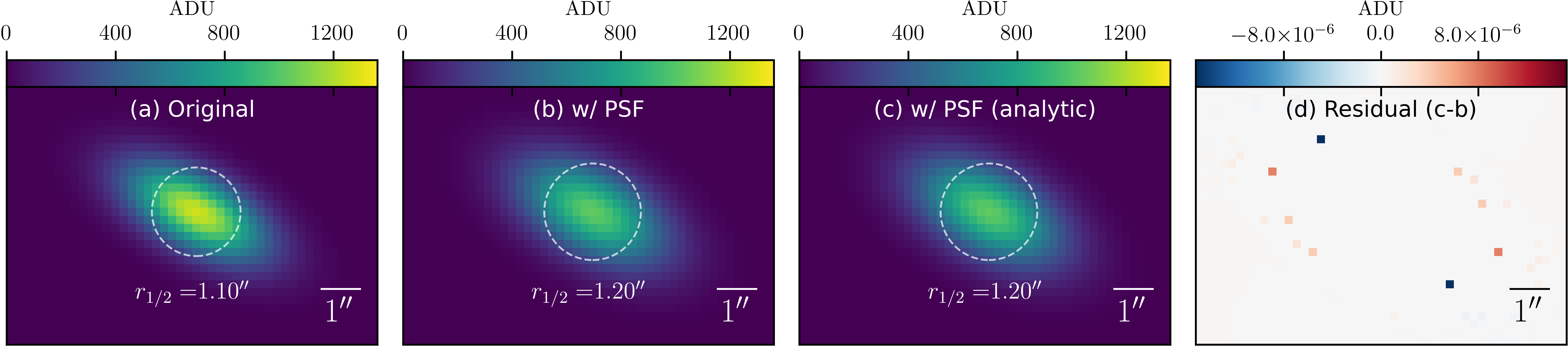}
        \caption{The image of a galaxy in the \texttt{BlendSim} pipeline (a) before PSF convolution, (b) after convolution with a circular PSF using \texttt{GalSim}, (c) after convolution with our analytical PSF model, and (d) the difference between the analytical approach (Section \ref{galprop}) and \texttt{GalSim}. The PSF FWHM is assumed to be $\sim 0.9~ \rm arcsec$. The calculated half-light radius, $r_{1/2}$, of each drawn galaxy is depicted by a dashed circle. The colour bars represent the co-added Analogue-to-Digital Unit (ADU) counts per pixel based on 1.3 yr worth of LSST observations. This unit test is a consistency check to verify our simple analytical approach used for PSF convolution in this study. The agreement between the two methods in this example is within $\sim 1.5 \times 10^{-5}$ ADU. }\label{fig:blendsimVSGalSim}
\end{figure*}

	\begin{figure}
	\centering
	\includegraphics[width=0.47\textwidth]{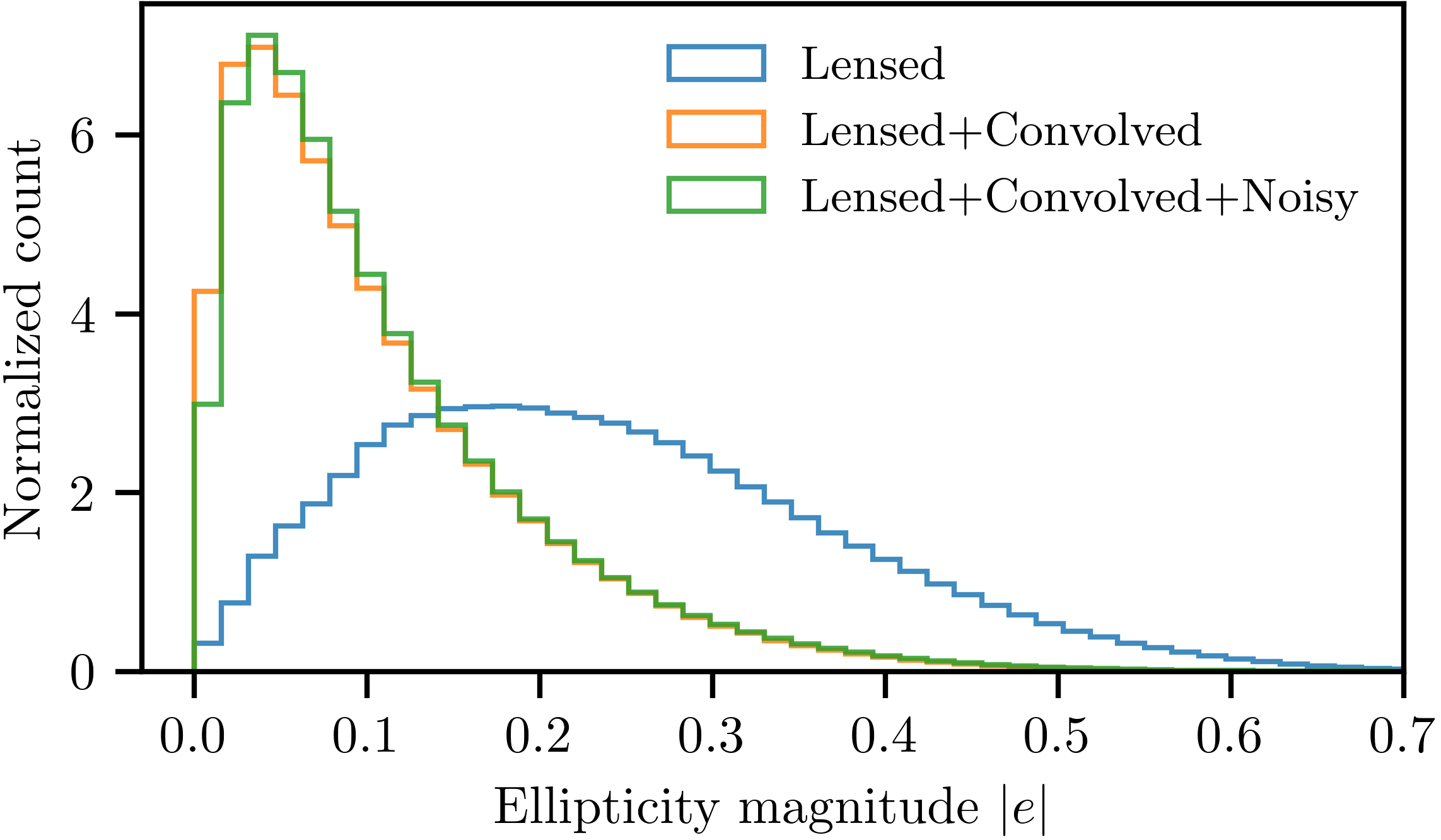}
	\caption{The ellipticity magnitude distribution of the galaxies in the \texttt{Buzzard} simulation. The blue histogram corresponds to lensed only shapes, orange to lensed and convolved with the PSF, and green to the noisy version of the lensed and convolved shapes. As expected, one can see the circularization effect of the PSF. The counts are normalized such that the integral over $e$ is unity. The addition of noise renders a very small fraction of the ellipticities (not shown here) useless by giving them ellipticity magnitudes above 1. They are excluded from our analysis. See Section \ref{galprop} for more details.}
	\label{edist} 
\end{figure}

The PSF convolution is done analytically in moment space. It is straightforward with known profiles, and also enables us to avoid memory issues that arise from the need for pixel-level convolution in large images. We model galaxies with Gaussian surface brightness profiles described by
\begin{equation}
G(x) = A  \exp\big[-\frac{1}{2}(x-\mu)^{T} \Sigma^{-1}  (x-\mu)\big],
\label{gaussgal}
\end{equation}
with $\mu$ being the position of the galaxy centre, $A$ the normalized flux of the galaxy, and $\Sigma$ the shape covariance matrix.
Our analytical method of PSF convolution requires the shape covariance matrix of the galaxies (a.k.a. the second moments' matrix) in order to work in the Fourier domain. This covariance is a 2 by 2 matrix that encapsulates the ellipticity and the position angle of each galaxy as seen on the 2D sky. Assuming a simple Gaussian model of equation  \eqref{gaussgal} for both the galaxy and the PSF, we add the second moments of the galaxy ($Q^{\prime}_{ij}$) and the PSF ($P_{ij}$) to get the combined moments of
\begin{equation}\label{qconv}
    Q_{ij} = Q^{\prime}_{ij}+P_{ij},
\end{equation}
where $i, j \in \{1,2\}$, and convert them back to ellipticity according to the definition 
\begin{equation}\label{edefinition}
    e = \frac{ Q_{11}-Q_{22} + 2 i Q_{12}}{Q_{11}+Q_{22}+2\sqrt { Q_{11} Q_{22}-Q_{12}^2} }.
\end{equation}
An area measure is to be obtained from the trace of the second moments' matrix as $T=Q_{11}+Q_{22}$. Using the same Gaussian assumption, one can find the adaptive moments' sigma for a circular Gaussian as $\sigma_{\rm Q}=(\det Q)^{1/4}$ where
\begin{equation*}
Q=\begin{pmatrix} Q_{11} & Q_{12} \\ Q_{21} & Q_{22} \end{pmatrix}.
\end{equation*}

An illustration of the shape circularization and galaxy size boost that occur when the true galaxy image is analytically convolved with the PSF that we adopt are shown in Fig. \ref{fig:blendsimVSGalSim}, and is compared with the pixel-level convolution of \texttt{GalSim} \citep{GalSim}. It shows good agreement between our analysis (Fig. \ref{fig:blendsimVSGalSim}c) and the \texttt{GalSim} pixel-based output (Fig. \ref{fig:blendsimVSGalSim}b). Convolution with the PSF is assumed to preserve the flux thereby not changing the magnitudes. Moreover, Fig. \ref{edist} shows how PSF convolution affects the observed ellipticity distribution of galaxies by reducing their ellipticity magnitudes.

We generate the simulated LSST-like magnitudes in 6 optical bands ($ugrizy$, covering the wavelength range 320--1050 nm) from \texttt{Buzzard} and estimate their errors utilizing and modifying the Dark Energy Science Collaboration (DESC) Photo-$z$ Working Group Simulations Generator\footnote{\url{https://github.com/LSSTDESC/PhotoZDC1}}. This implements the simple magnitude error model described in \cite{Ive19}. In consideration of the fact that galaxies are extended sources, we add the FWHM of each galaxy's light profile in quadrature to the seeing FWHM ($\theta_{\rm eff}$) when applying equation 6 of \cite{Ive19}. We assign magnitude errors in each of the six bands assuming the equivalent of 1.3 years of LSST observations, consistent with the depth of our sample. This number of years translates into $\rm S/N>20$ for all the point sources brighter than $i=24$.
We also take into account the presence of unrecognized blends by re-measuring observed magnitudes and their errors after the blending emulation (outlined in Section \ref{blendemu}), i.e. by adding observational noise to the combined (noise-free) magnitude of each unrecognized blend.

\begin{figure}
	\centering
	\includegraphics[width=0.48\textwidth]{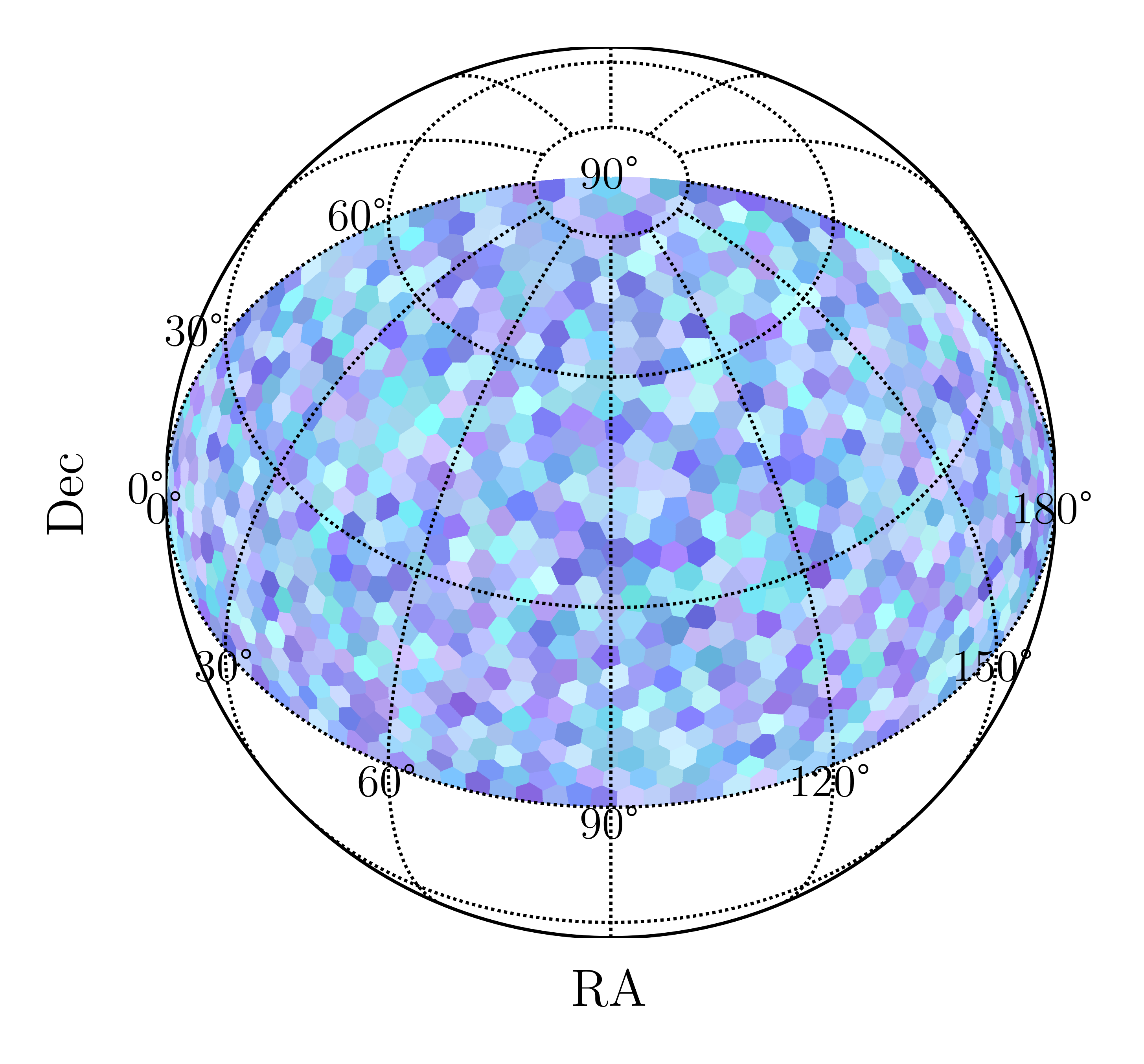}
	\caption{The part of the sky that is covered by the sample from each realization of the \texttt{Buzzard} simulation. The initial footprint is defined prior to lensing being applied. We avoid abnormal overdense and underdense regions along the edge of \texttt{Buzzard}'s footprint by removing areas within 0.2 degrees of the edge. The different alternating shades of blue represent 1000 different jackknife regions used in the covariance matrix calculation (Section \ref{covariance}).}
	\label{mysample} 
\end{figure}

Fig. \ref{mysample} shows the coverage of our galaxy sample on the sky.  Using the spectroscopic and photometric redshift data, we split the sample into four tomographic redshift bins (0.3,0.5), (0.5,0.7), (0.7,0.9), and (0.9,1.2). We exclude low redshift galaxies where the lensing signal is low. We also chose a larger tomographic bin width ($\Delta z_{\rm tomo} = 0.3$ as opposed to 0.2 used in the other bins) for the highest redshift bin to improve the number count for that bin. The maximum redshift of $z=1.2$ we adopted here is the same as the analysis setup of LSST DESC Science Requirement Document \citep{srd} for the large-scale structure. However, they do not have an upper redshift cutoff for their ``source'' sample. See Section \ref{photozest} for information regarding our upper redshift cutoff and photometric redshift estimation. 

To model more realistic measured ellipticities rather than adopting the noise-free (only lensed and PSF-convolved) ellipticities, we include an analytical formulation of the effects of pixel noise. We use the analytical model proposed by \cite{epsnoise} which captures the entire process from noisy images to shear estimates. One advantage of this model is that it only requires the signal-to-noise ratio of each observed galaxy in order to predict the observed ellipticities from noise-free ellipticities. We use this model to add simulated pixel noise to the PSF-convolved shapes based on the expected $i$-band flux signal-to-noise (S/N) level derived from the magnitude errors, ${\rm d}m_{\rm i}$:
\begin{equation}
    \frac{S}{N} = \frac{1}{10^{{\rm d} m_{\rm i}/2.5} - 1}.
\end{equation}
We show the resulting ellipticity distribution in Fig. \ref{edist} (green histogram). In the same figure, the histogram of the noise-free lensed shapes along with their PSF-convolved version are plotted for comparison. Due to the effect of noise on moment-based shapes, a minuscule fraction of the ellipticity magnitudes got pushed to values above unity and will not be used in our analysis. To model measured galaxy sizes rather than adopting the true lensed-only catalogue sizes, we include an analytical formulation of the observed galaxy sizes due to signal-to-noise ratio and the ellipticity magnitude. We employ an analytical model \citep[Table 1]{refregier12} to add Gaussian noise of standard deviation
\begin{equation}
\sigma_{r_{1/2}} = r_{1/2} \sqrt{\frac{1+\epsilon^2}{2(S/N)^2}}, ~~ \epsilon = \frac{2|e|}{1+e^2}
\end{equation}
to the half-light radii $r_{1/2}$ of our galaxies as a function of their ellipticity magnitude and ($i$-band) flux S/N, assuming the relative error in the half-light radius is the same as the relative error in the radius mean $a=\sqrt{a_1^2+a_2^2}$ with $a_1$ and $a_2$ being the semimajor and semiminor axes. The ellipticity, $e$, is defined in equation \eqref{edefinition}.

\section{Blending}\label{blending}
\subsection{Importance of blending in weak lensing}
\begin{figure}
	\centering
    \includegraphics[width=1\linewidth]{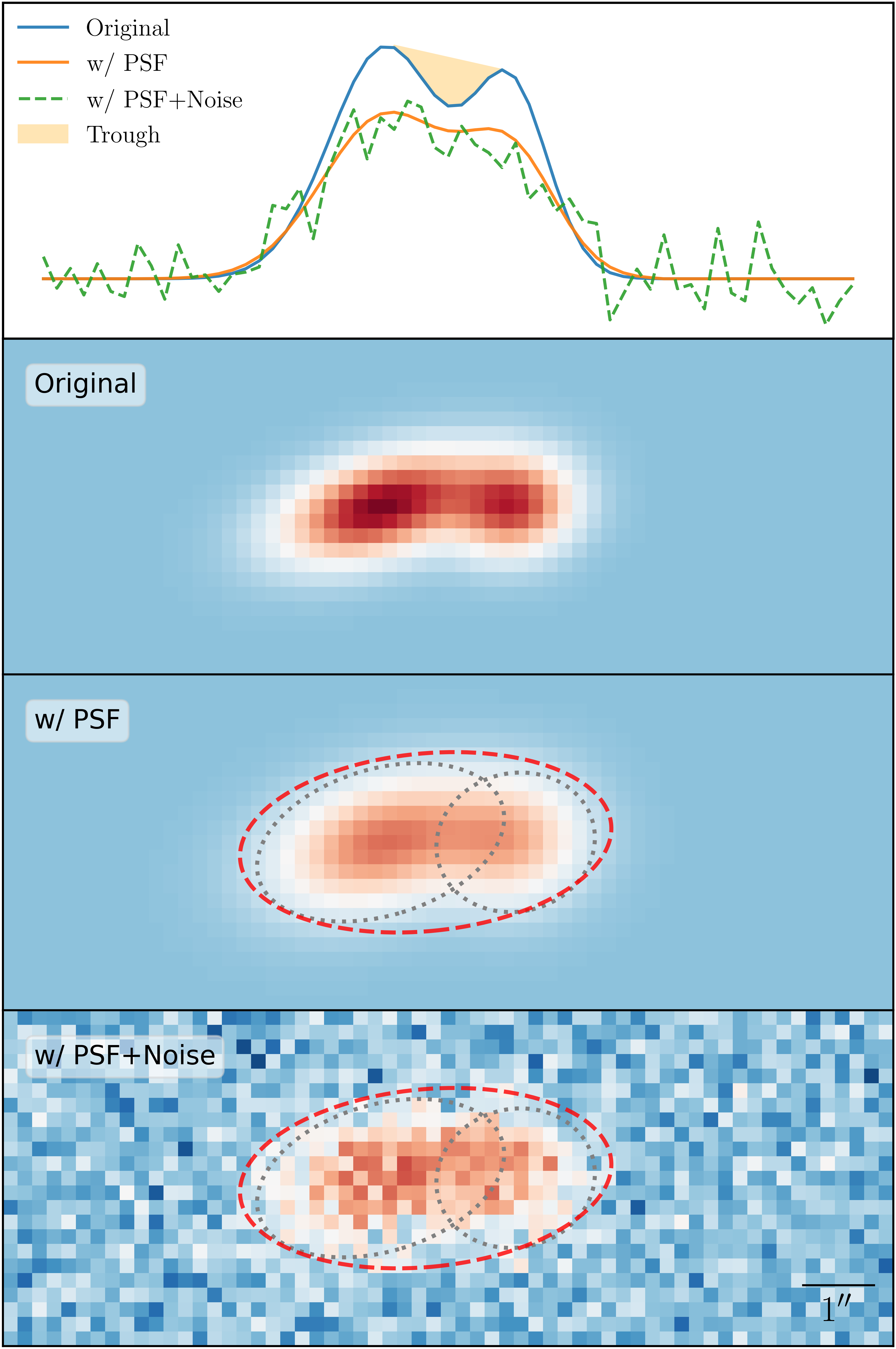}
	\caption{Synthetic images of two overlapping galaxies (\textit{bottom} three panels), simulated using \texttt{GalSim}, alongside their corresponding surface brightness profiles across the blend (\textit{top} panel) showing how unrecognized blends cause issues with galaxy detection in blended profiles.  The two \textit{middle} panels are noise-free, while the \textit{bottom} panel has Poisson noise as a result of the flux of galaxies and a sky level of $i_{\rm sky}=20.48 \rm ~mag/arcsec^2$. The $i$-band magnitudes of the two galaxies are 22.60 and 23.35, as viewed from left to right. The PSF FWHM of $\sim 0.9~ \rm arcsec$ and 1.3 yr worth of LSST observations are assumed. The amplitude of the trough between the two peaks, the PSF, and the sky level are the key factors in determining whether deblending algorithms can recognize the blend or not. The grey dotted ellipses indicate the shapes of individual galaxies after being convolved with the PSF. The red dashed ellipses is the analytical shape computed from the combined moments of (PSF-convolved) galaxies in the blend (equation \ref{Qij}). }
	\label{gblend}
\end{figure}

Blending of galaxy surface brightness profiles characteristically can be divided into two categories: (1) recognized blending, in which multiple surface brightness peaks can be recognized in a blended image, and (2) unrecognized or ambiguous blending, where the deblending algorithm is unable to deduce more than one surface brightness peak (see Fig. \ref{gblend}). We will focus here more on this latter aspect.

As noted by \citet{rachel}, the two prevalent systematic errors in weak lensing are thought to be those of blending on shear measurement and on photometric redshift.  In an unrecognized blend, the single object can often have observed shapes and fluxes that are quite different in value from those of any of the two or more constituent objects that comprise the blend.  
We observe in our data set that most of the unrecognized blends occur between objects at different redshifts, which agrees with the findings of Kirkby et al. (in preparation). Therefore, for most blends, it is not clear what to make of the collective shear and photometric redshift estimates. A recent study by \cite{maccrann2022} looked at the joint effects of blending in both shear and photometric redshift calibration using image simulations in the context of DES Year 3.

A significant number of objects will be part of unrecognized blends upon observation in the LSST survey. Removing (unrecognizably) blended objects from the data set is not an option, as they cannot be isolated and removed if they are unrecognized, by definition.  \citet{dawson2016} estimates the unrecognized blend fraction to be $\sim$14 per cent for the LSST gold sample. \cite{sanchez21} revisit the statistics of the blends.
\cite{chang13} present the impact of rejecting blend systems that exhibit major blending when determining expected galaxy source number density and redshift distribution from LSST. The rejection of these recognized blends and their impact on cosmology is the focus of Section \ref{recbl}.

In weak lensing measurements, unrecognized blending poses both a noise problem and a bias problem: the random nature of overlapping galaxies can add spurious ellipticity noise, and the corresponding random ellipticities dilute the shear signal causing a shear polarizability bias.  
It has been shown by \citet{dawson2016} that as a result, shear variance will increase both through a reduction in counts and a higher contribution to shape noise than non-blended galaxies. However, multiplicative shear bias is of greater significance here.

\subsection{Emulating blending phenomenon in Buzzard}\label{blendemu}
Our goal is to emulate blending effects in a sample analogous to what will be used in shear measurements in the upcoming deep imaging surveys such as LSST, sometimes referred to as the ``gold sample'' and often defined by an apparent magnitude cut, as we previously discussed throughout Section \ref{bz}.  However, we cannot simply limit ourselves to galaxies brighter than some limiting magnitude for blending emulation, since the steep slope of the apparent magnitude number counts observed in survey data indicate that galaxies with fluxes up to two magnitudes fainter will contribute enough flux to bias shape and photo-$z$ measurements if blended with an object from our weak lensing (i.e., source) sample.  Therefore, we must include galaxies fainter than our fixed flux limit when emulating blends.  In the same vein as \citet{dawson2016}, we include and examine galaxies up to two magnitudes fainter (to $i<26$) that will contribute to unrecognized blends in our resultant cosmology sample ($i<24$) during emulation. Note, however, that additional cuts are made to define the final source sample analyzed in this study. These cuts are described in Section \ref{cosmicshearprobe}.

Assuming a simple Gaussian galaxy model of equation \eqref{gaussgal}, the quadrupole moments of the combined shape of $n$ galaxies that are blended together become (cf. \citealt[equation~B3]{dawson2016})

\begin{equation}
    Q^{\rm blend}_{ij} = \frac{1}{\sum_{k=1}^{n} A^{(k)}} \sum_{k=1}^{n} A^{(k)} \big[(\mu_{i}^{(k)}-c_i) (\mu_{j}^{(k)}-c_j) + \Sigma_{ij}^{(k)} \big],
\label{Qij}
\end{equation}
where $k$ loops over the galaxies in the blend, $c_i$ and $c_j$ are the coordinates of the luminosity centroid of the blend, and $\Sigma_{ij}^{(k)}$ is the element $ij$ of the shape covariance matrix of the $k$th galaxy in the blend.
The ellipticity of the blend can then be determined by using equation \eqref{edefinition}.

To emulate the effects of unrecognized blending on the truth catalogue, we first identify potential blend candidates in crowded regions (a.k.a. blended scenes), defined as two or more galaxies closer than the spatial separation $\theta_{\rm b}=4.0$~ \rm arcsec\,measured from galaxy centroids\footnote{For efficiency, this process is performed by {\sc healpix}, and therefore does not include the pairs that cross the {\sc healpix} borders. We believe that disregarding such pairs does not change our overall conclusion, especially since the {\sc healpix} pixels are relatively large ($\sim$53.7 square degrees).} (the validity of this choice is explained later in this section). For this purpose, we use the Friends-of-Friends (FoF) grouping algorithm from the \texttt{SkyLink}\footnote{\url{https://github.com/enourbakhsh/SkyLink}} package which enhances efficiency by using $k$-dimensional trees \citep{kdtree} and graphs\footnote{By definition, a galaxy can only belong to a single FoF group or blended scene, i.e., in the event of an intersection between FoF groups, the groups will merge.}.  We then inspect galaxies in the blended scenes to confirm whether they are blends or not. The inspection is based on simulated image cutouts of the blended scenes generated in real space using \texttt{GalSim}. The images do not undergo pixel-level convolution due to galaxies already being PSF-convolved prior to image simulation, as detailed in Section \ref{galprop}. All simulated images are based on 1.3 yr worth of LSST observations with the parameters taken from table 2 of \cite{Ive19}.  The source detection is done by taking advantage of the software package \texttt{SEP} \citep{barbary16}, an in-memory {\sc python} implementation of the \texttt{Source Extractor} or \texttt{SExtractor} software \citep{bertin96}. The detection process does not estimate the background level locally; instead, it uses the global sky level utilized by \texttt{GalSim} to create images, i.e. $i_{\rm sky}=20.48 \rm ~mag/arcsec^2$. In cases where detection and truth do not correspond, we treat all non-blended galaxies in each detected segment as a single combined galaxy (i.e. an unrecognized blend) and create a so-called ``blended'' catalogue by using their collective values to modify the non-blended catalogue. We sum non-blended lensed-only galaxy fluxes within each unrecognized blend to get the total flux, convert it back to the total magnitude, and add noise to the magnitude, as done in Section \ref{galprop} for galaxies in the non-blended catalogue. We calculate the combined second moments, $Q^{\rm blend}$, using equation \eqref{Qij}. The elements of this matrix give us the half-light radius of each unrecognized blend. They can also be plugged into equation \eqref{edefinition} to yield the collective shape for each blended system. We then add noise to the half-light radii and shapes of the unrecognized blends in accordance with the methods described in Section \ref{galprop}. 
If a galaxy in the non-blended catalogue goes undetected by \texttt{SEP}, we: (1) keep it in our blended catalogue as an isolated galaxy if the observed (noisy) version of its magnitude is brighter than our sample cut at $i=24$, or (2) remove it from our blended catalogue if it is fainter than $i=24$. However, such non-detections tend to be quite faint in general and thus are more likely to fall under the latter category.

\begin{figure}
	\centering
   \includegraphics[width=1
   \linewidth]{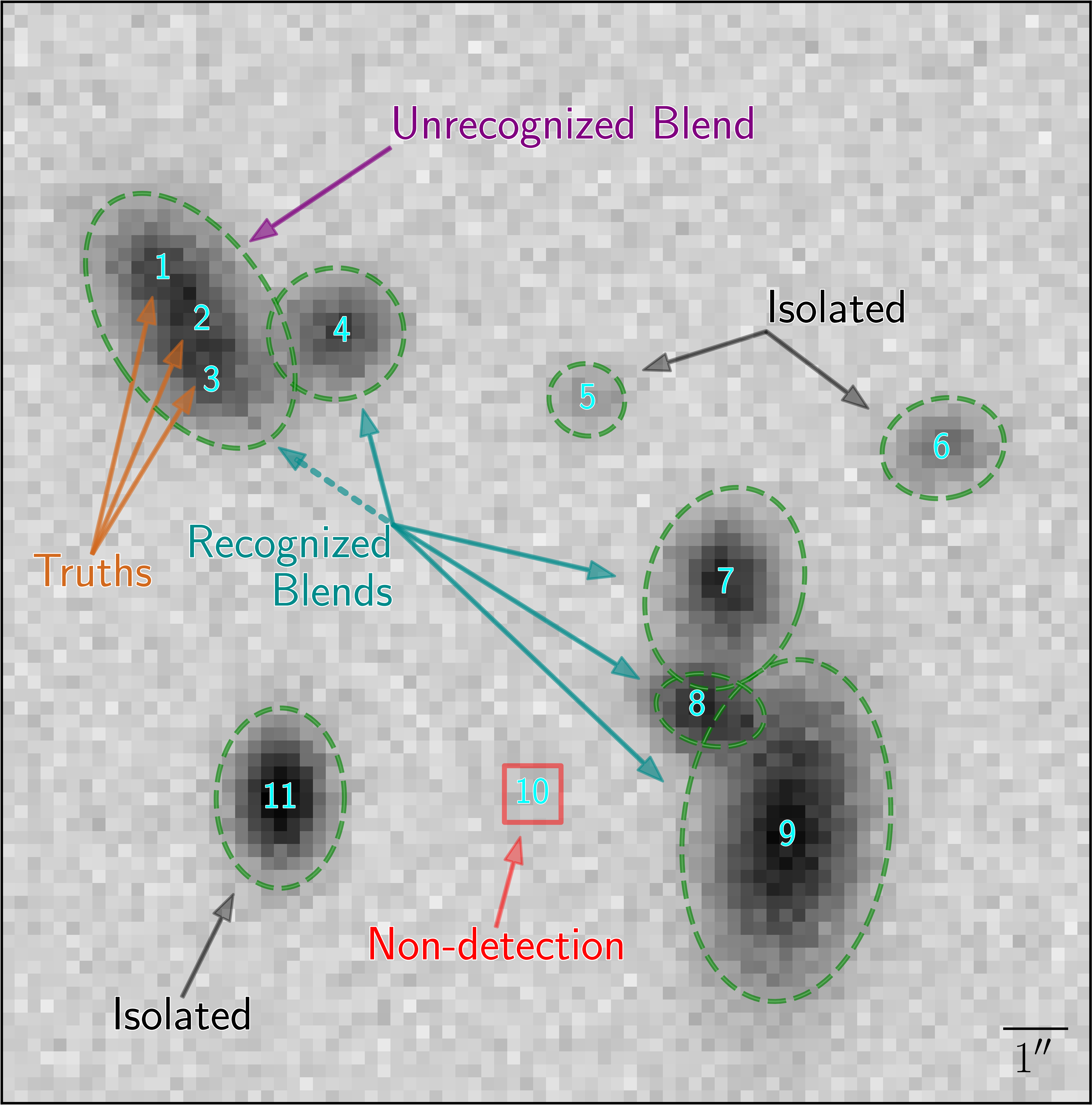}
	\caption{An illustrative image cutout of a crowded region. Detected sources are indicated by green dashed ellipses and the positions of true sources from the input catalogue by cyan numbers. A detection may be an unrecognized blend (galaxies 1, 2, and 3, collectively), a recognized blend (galaxies 4, 7, 8, and 9), or an isolated source (galaxies 5, 6, and 11). The red square (around galaxy 10) represents a non-detected object that exists in the truth catalogue but is not detected by the source detection algorithm. The unrecognized blend in the example above is recognizably blended with galaxy 4 and is therefore viewed as a recognized blend by an observer unaware of the truth. This case is outlined with a `dashed' arrow to inform the reader that it requires special attention. We use \texttt{GalSim} and \texttt{SExtractor} to create this image.}
	\label{blendpic} 
\end{figure}

\begin{figure*}
	\centering
	\begin{subfigure}[h]{0.48\textwidth}
        \includegraphics[width=1\linewidth]{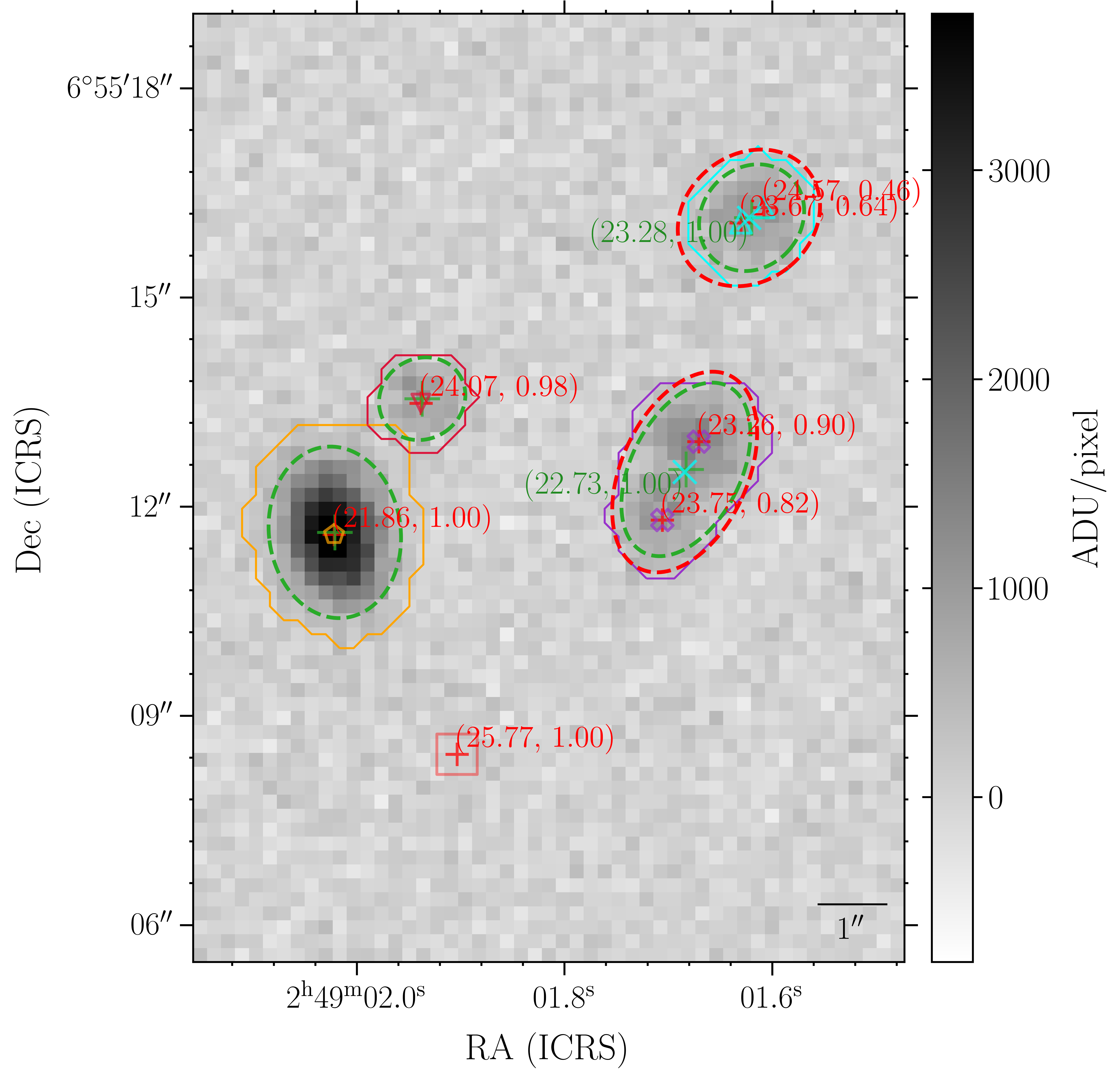}
        \label{sep1a} 
    \end{subfigure}
    \qquad
	\begin{subfigure}[h]{0.48\textwidth}
        \includegraphics[width=1\textwidth]{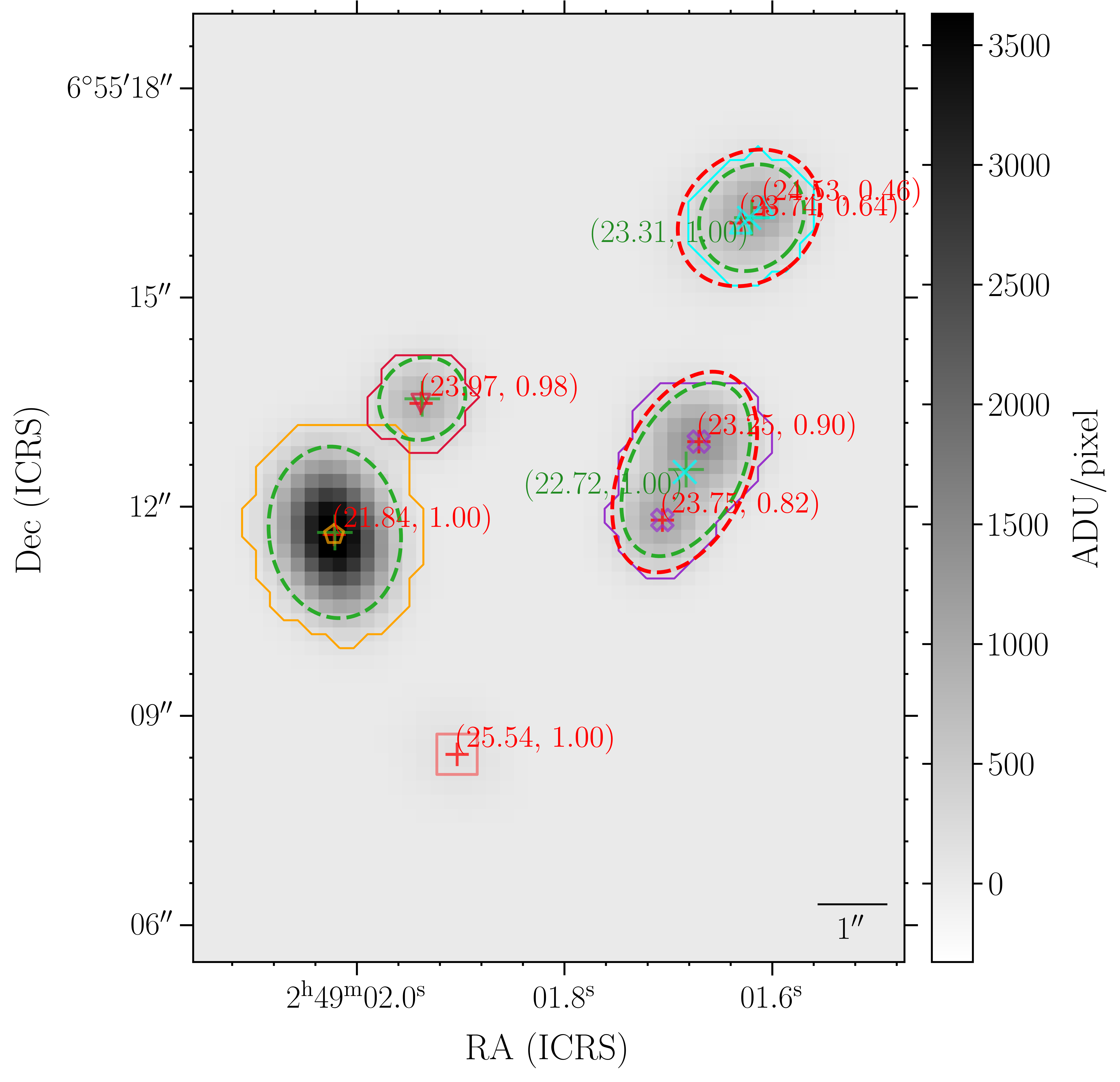}
        \label{sep1b} 
\end{subfigure}
    
	\caption{An example of the performance of the detection and deblending module in \texttt{BlendSim}. This module employs \texttt{SExtractor} through \texttt{SEP}.. All the galaxies from the non-blended catalogue are shown by the red pluses. We use different markers for galaxies that belong to different detections according to the segmentation map. Two unrecognized blends each consisting of two objects are marked by cyan triangles and orchid hollow X shapes, while two isolated galaxies are marked by a crimson triangle and a yellow pentagon. The values in the parentheses are the $i$-band magnitude and purity (Section \ref{recbl}), respectively (in red for individual galaxies from the non-blended catalogue, and in green for blends collectively). A galaxy at the bottom which is undetected by \texttt{SExtractor} is marked by a faint red square, and, in this case, is removed from the cosmology sample since it is fainter than $i=24$.  The segments defined for each detection are overlaid using the same colour as the markers. The {\it left-hand} panel, which is used for detection, has Poisson noise due to the flux of the galaxies and a sky level of $i_{\rm sky}=20.48 \rm ~mag/arcsec^2$, while the {\it right-hand} panel is noiseless.
	The green dashed ellipses with green pluses at their centres are the shapes measured by \texttt{SExtractor}, and the red dashed ellipses on the unrecognized blends with cyan crosses at their centres (marking the analytical luminosity centre of the blend) are the shapes computed analytically from the combined moments of individual galaxies in the blend without pixel noise. We consider the noisy version (method described in Section \ref{galprop}) of the latter as the unrecognized blends' shapes in calculating correlations for the blended source sample. The colour bars depict the co-added pixel values over 1.3 yr of LSST.}
	\label{sep1}
\end{figure*}

The emulation has resulted in approximately 12 per cent of galaxies in the blended catalogue being unrecognized blends. See Figs \ref{blendpic} and \ref{sep1} for a simplified toy example and real production example of the image simulation, respectively. The detection and deblending parameters that we specified for \texttt{SExtractor} through \texttt{SEP} are listed in Table \ref{sepparams}.
Later in Section \ref{cosmicshearprobe}, we calculate galaxy angular correlations and check for any systematic changes that occur due to blending by comparing measurements made using our blended catalogue to measurements made on the equivalent \texttt{Buzzard} catalogue where blending effects were not emulated. 
 
\begin{table}
    \centering
    \begin{tabular}[t]{lr}
    \toprule
    {\bf Parameter} & {\bf Value}\\
    \midrule
    {\tt DETECT\_NTHRESH} & 2.0 \\
    {\tt DETECT\_MINAREA} & 5 \\
    {\tt DEBLEND\_NTHRESH} & 32 \\
    {\tt DEBLEND\_MINCONT} & 0.0001 \\
    {\tt FILTER} & gauss\_4.0\_7x7.conv \\
    {\tt CLEAN\_PARAM} & 1.1 \\
    \bottomrule
    \end{tabular}
    \caption{Specified \texttt{SExtractor} detection and de-blending parameters through the {\sc python} library for Source Extraction and Photometry (\texttt{SEP}).}
    \label{sepparams}
\end{table}

\begin{figure}
	\centering
   \includegraphics[width=\linewidth]{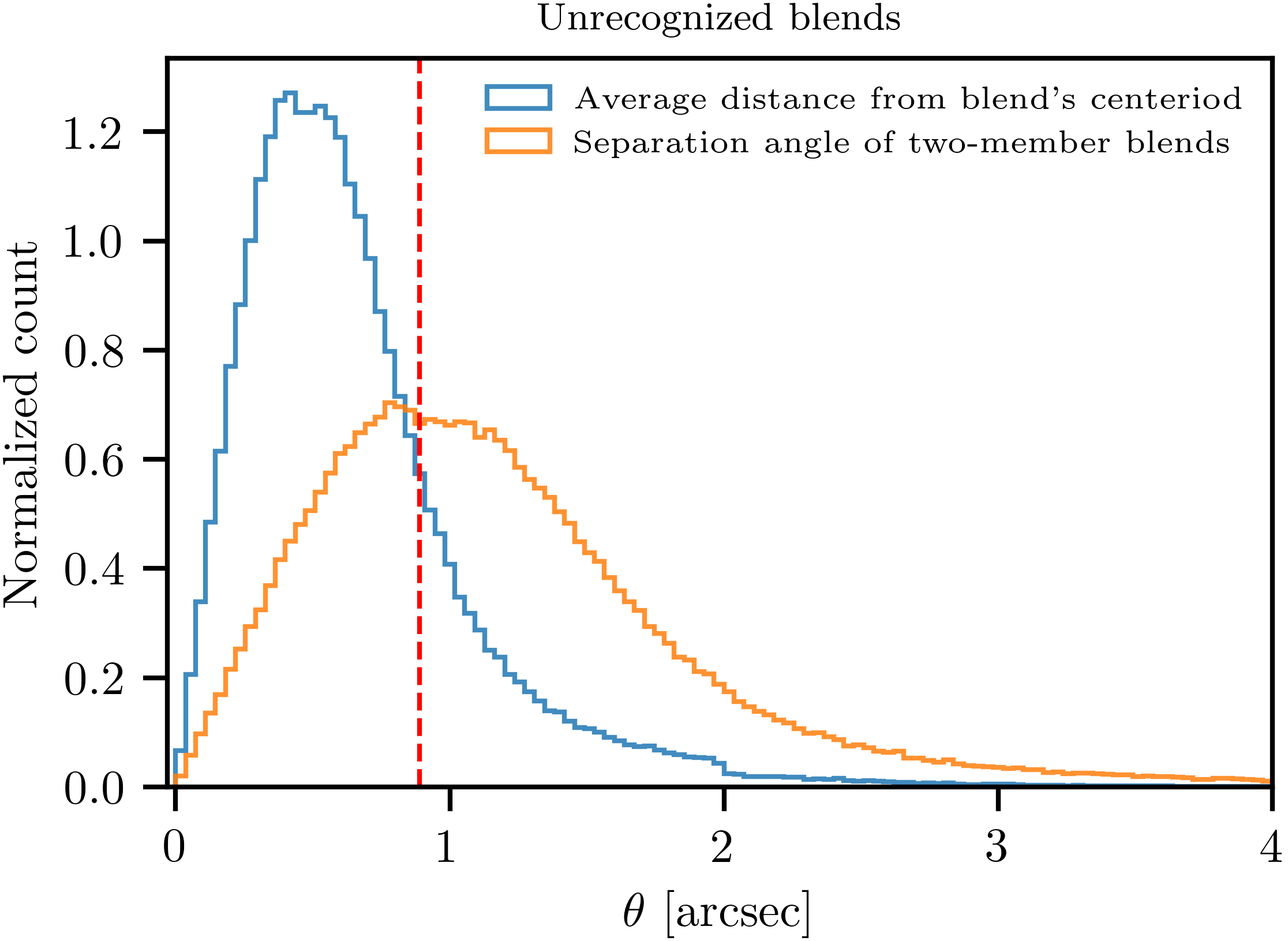}
	\caption{The distribution of the average distance from the unrecognized blend's centroid (blue) and the separation angle of two-member unrecognized blends (orange), obtained from one {\sc healpix} cell of the blended cosmology sample. Note that the orange histogram peaks at about the PSF FWHM of our sample marked by the red dashed line at $\sim 0.9 ~ \rm arcsec$.  The blue histogram uses the luminosity centroid as a reference for blends with two or more members and peaks at $\sim 0.45 ~ \rm arcsec$ which is half of the PSF FWHM. The very small tail at $\theta=4 ~ \rm arcsec$ indicates that using this rather conservative value as a separation angle threshold when defining blend candidate groups in Section \ref{blendemu} is appropriate.}
	\label{sepangpics}
\end{figure}

Fig. \ref{sepangpics} shows the distribution of the angular separation between the confirmed unrecognized blend components for pairs (orange histogram) and for all unrecognized blends, including multiples (blue histogram). For the multiples, we use the average angular distance of the blend components from the luminosity centroid of the blend. There is still a very small tail at $4~ \rm arcsec$ in the orange histogram (the equivalent of $2~ \rm arcsec$ for the blue histogram). Thus, our conservative choice of the separation angle threshold ($\theta_{\rm b}=4 ~ \rm arcsec$) for finding the blended scenes is appropriate.

A measure of purity -- described in detail in Section \ref{recbl} -- can be used to quantify the degree of blending with known surface brightness profiles. Therefore, during the emulation process, we calculate the purity value for every unrecognized blend, recognized blend, or otherwise isolated galaxy in the blended scenes (all brighter than $i=24$) to be used later in Section \ref{recbl} for excluding severe cases of recognized blends. A relevant point to mention is that unrecognized blends themselves can be (recognizably) blended with other galaxies in which case they will be regarded as recognized blends by the observer who has no insight into the truth. This is exemplified by the unrecognized blend in Fig. \ref{blendpic}.

\section{Photometric RedShift Estimation}\label{photozest}
\begin{figure*}
	\centering
	\begin{subfigure}[h]{0.47\textwidth}
        \includegraphics[width=1\linewidth]{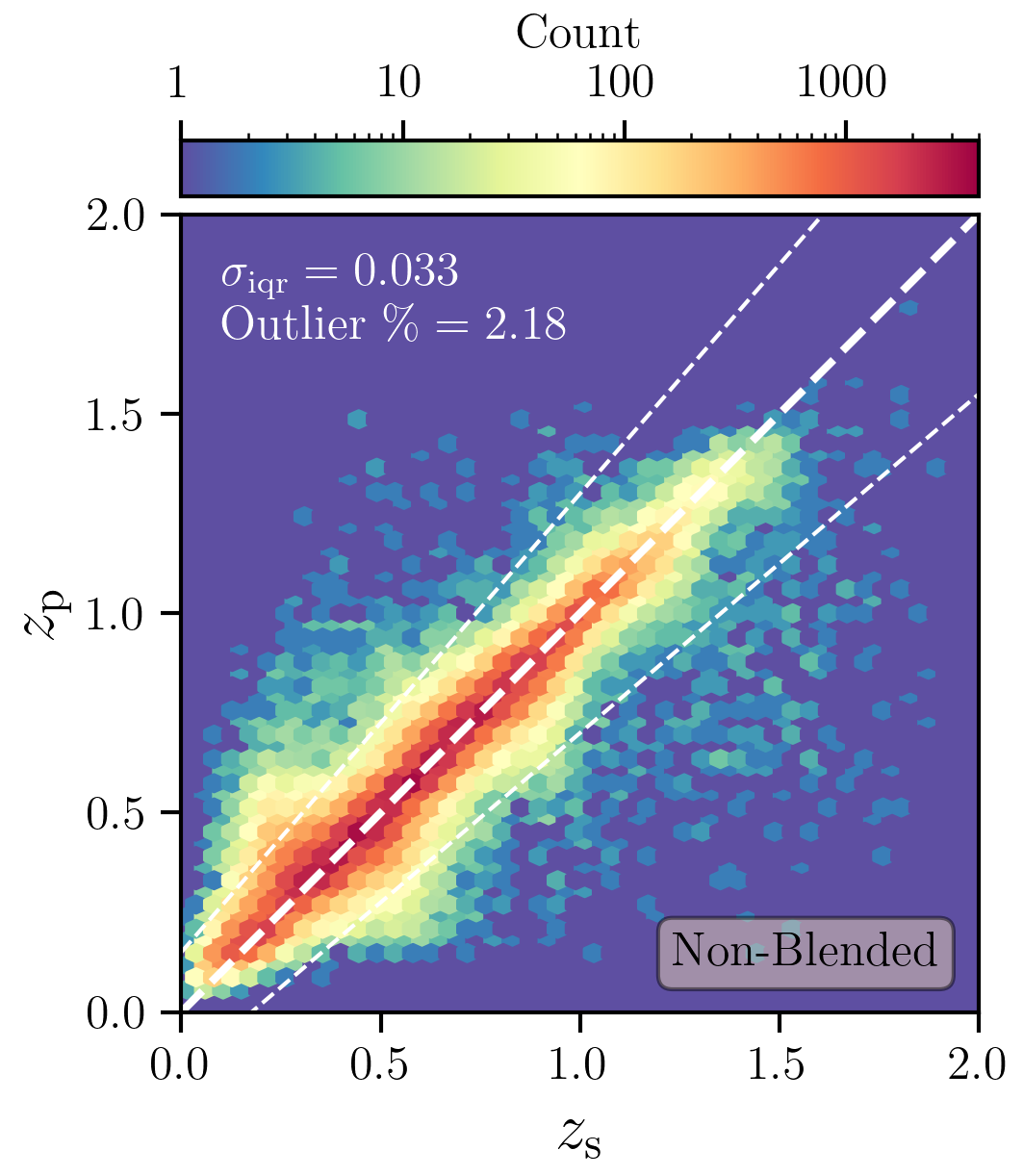}
        \label{zzplot_nb} 
    \end{subfigure}
    \qquad
	\begin{subfigure}[h]{0.47\textwidth}
        \includegraphics[width=1\textwidth]{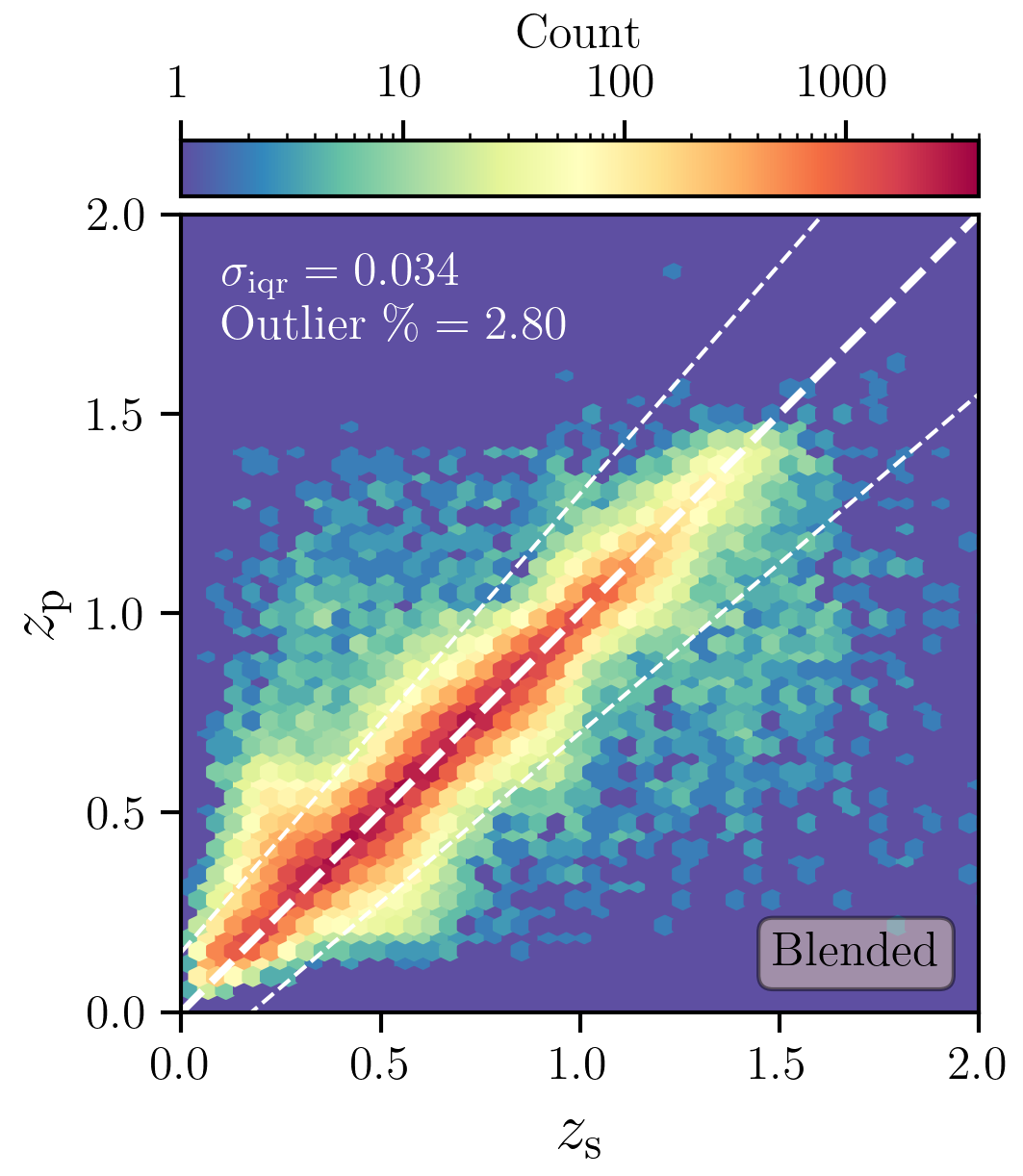}
        \label{zzplot_b} 
    \end{subfigure}
    
	\caption{Photometric redshift ($z_{\rm p}$) point estimate versus the spectroscopic redshift ($z_{\rm s}$) for $\sim 200,000$ randomly selected galaxies in each of the non-blended ({\it left-hand} panel) and blended ({\it right-hand} panel) source samples, i.e. with $22<i<24$ and lensing quality cuts described in Section \ref{cosmicshearprobe} as the validation sets drawn from \texttt{Buzzard}. We employed a hexagonal binning scheme where the value of the hexagon is determined by the number of points in the hexagon mapped onto the colour bars shown at the top of each panel. The central dashed line represents a one-to-one match between $z_{\rm p}$ and $z_{\rm s}$ and the two dashed lines around the central line draw the borders for the catastrophic outliers where $|(z_{\rm p}-z_{\rm s})/(1+z_{\rm s})|>0.15$. The plots are annotated by the values for the interquartile range $\sigma_{\rm iqr}$ in $e_{z}=(z_{\rm p}-z_{\rm s})/(1+z_{\rm s})$ and the percentage for the catastrophic outliers. The blended set shows a larger interquartile range and outlier rate. Note that the $z_{\rm s}$ values for the unrecognized blends in the blended set are taken from the redshift of the brightest galaxy in the $i$ band among the galaxies in the blend.}
	\label{zzplot}
\end{figure*}

We use Trees for Photo-Z (TPZ\footnote{\url{https://github.com/mgckind/MLZ}}; \citealt{tpzpaper}), to estimate photometric redshifts for our galaxies. TPZ is a supervised machine learning algorithm in {\sc python} that uses prediction trees and random forest techniques to produce both robust photometric redshift PDFs and ancillary information for a galaxy sample. \cite{bh10} and \cite{newman15} showed that $\sim 10^4$ spectra are necessary to calibrate photo-$z$'s to Stage IV requirements of the LSST Science Requirements Document (SRD)\footnote{Obtainable from \url{https://docushare.lsstcorp.org/docushare/dsweb/Get/LPM-17}}. With these findings in mind, we train our estimator consisting of 100 trees using a training set of $10^5$ galaxies with spec-$z$'s randomly drawn from our cosmology sample. Fig. \ref{zzplot} provides an illustration of the algorithm's performance for point estimates for the blended and non-blended validation samples. The spectroscopic redshift and photometric redshift point estimates are denoted by $z_{\rm s}$ and $z_{\rm p}$, respectively. We have restricted our redshift tomography to $z<1.2$ beyond which the reliability of photo-$z$ estimation is expected to decline significantly.
There is a slight tail due to the outliers at low $z_{\rm s}$ in the blended case because $z_{\rm s}$ is taken from the redshift of the brightest galaxy (in the $i$ band), which tends to be at lower redshift than other galaxies in the blend.

Figs \ref{dndz_s_specz}, \ref{dndz_s_photoz}, and \ref{dndz_s_photoz_pdf} show our tomographic redshift binning with spec-$z$ and photo-$z$ for the non-blended source sample. While one can see a good agreement between the two point estimates (Figs \ref{dndz_s_specz} and \ref{dndz_s_photoz}), the photo-$z$ PDFs (Fig. \ref{dndz_s_photoz_pdf}) have a smooth and long tail in the distribution due to relying on the photometric data to estimate the redshift probabilistically. To continue our analysis for the spec-$z$ tomography in the next section, we make the same assumption as in Fig. \ref{zzplot}, by considering the spectroscopic redshift of an unrecognized blend to be the spectroscopic redshift of the brightest galaxy that went into the blend. However, we have no need to make such an assumption for the photo-$z$'s where we use the collective photometric data of the unrecognized blend to determine its redshift point estimate and PDF.  With regard to the photo-$z$ tomography, we place galaxies into various redshift bins according to their photo-$z$ point estimates, but we use their stacked photo-$z$ PDFs for the redshift number density, $n(z)$, throughout the paper.
	
The plots for the redshift distribution of the blended source sample (not shown) do not show any marked differences when compared to the non-blended source sample. This depends on what we are assuming about the impact of blending on each of the three measures of redshift tomography. 
Although a sizable proportion ($\sim$12 per cent) of the objects are blended and re-simulated, most blends have multiple faint objects blended with one bright source whose magnitude and colour dominates the blend thereby introducing a slight change in the observed photometry, and thus photo-$z$. To get more insight into the population of the faint neighbours, one useful quantity to look at could be the magnitude difference, $dm_{12}$, between the brightest and the second brightest galaxy inside each blend. We find that about 60 per cent of blends have $dm_{12} > 1.0$ and about 30 per cent of them have $dm_{12} > 2.0$. Hence, most galaxies inside unrecognized blends incorporate blended galaxies that are so faint that they do not affect the photometric redshift significantly.
	
\bigskip

\section{Weak lensing cosmological probe}\label{cosmicshearprobe}
Weak lensing enables the measurement of the mass power spectrum directly through cosmic shear, which refers to the distortion of the shapes of distant galaxies due to the gravitational effects of the intervening foreground large-scale mass structure. As a result, there are weak but statistically meaningful correlations in the way galaxies are aligned across the sky and in redshift due to the total mass distribution that their light rays encounter. In view of the small amplitude of the weak lensing effect, it is possible to identify these correlations between the shapes of galaxies by averaging over a sufficiently large sample. This can be done for different distances (i.e. tomographic redshift bins) that gives us a 3D map showing how dark matter is distributed in the universe, independent of assumptions about the relationship (i.e. galaxy bias) between visible/baryonic matter and dark matter. The angular shear--shear two-point correlation function can be formed as
\begin{equation}\label{eq:cosmicshear}
    \xi^{ij}_\pm(\theta) = \int_0^\infty\frac{d\ell\,\ell}{2\pi}\,C^{ij}_{\gamma \gamma}(\ell) J_{2\mp2}(\ell\theta),
\end{equation}
under the flat-sky approximation with $J_{\rm m}(x)$ being the $m$th order Bessel function of the first kind and $\theta$ the angular separation. The superscripts $i$ and $j$ label different tomographic redshift bins and $C^{ij}_{\gamma \gamma}$ represent the power spectra that contain information about the underlying dark matter. The correlation is computed as the sum  $\xi_{+}(\theta)$ and difference $\xi_{-}(\theta)$ of the product of tangential and cross components of shear in relation to the line connecting each pair of galaxies. Here, we add that $\xi^{ij}_\pm(\theta)=\xi^{ji}_\pm(\theta)$ for all tomographic bin pairs.

\begin{figure*}
	\centering
	\begin{subfigure}[a]{0.59\textwidth}
        \includegraphics[width=1\linewidth]{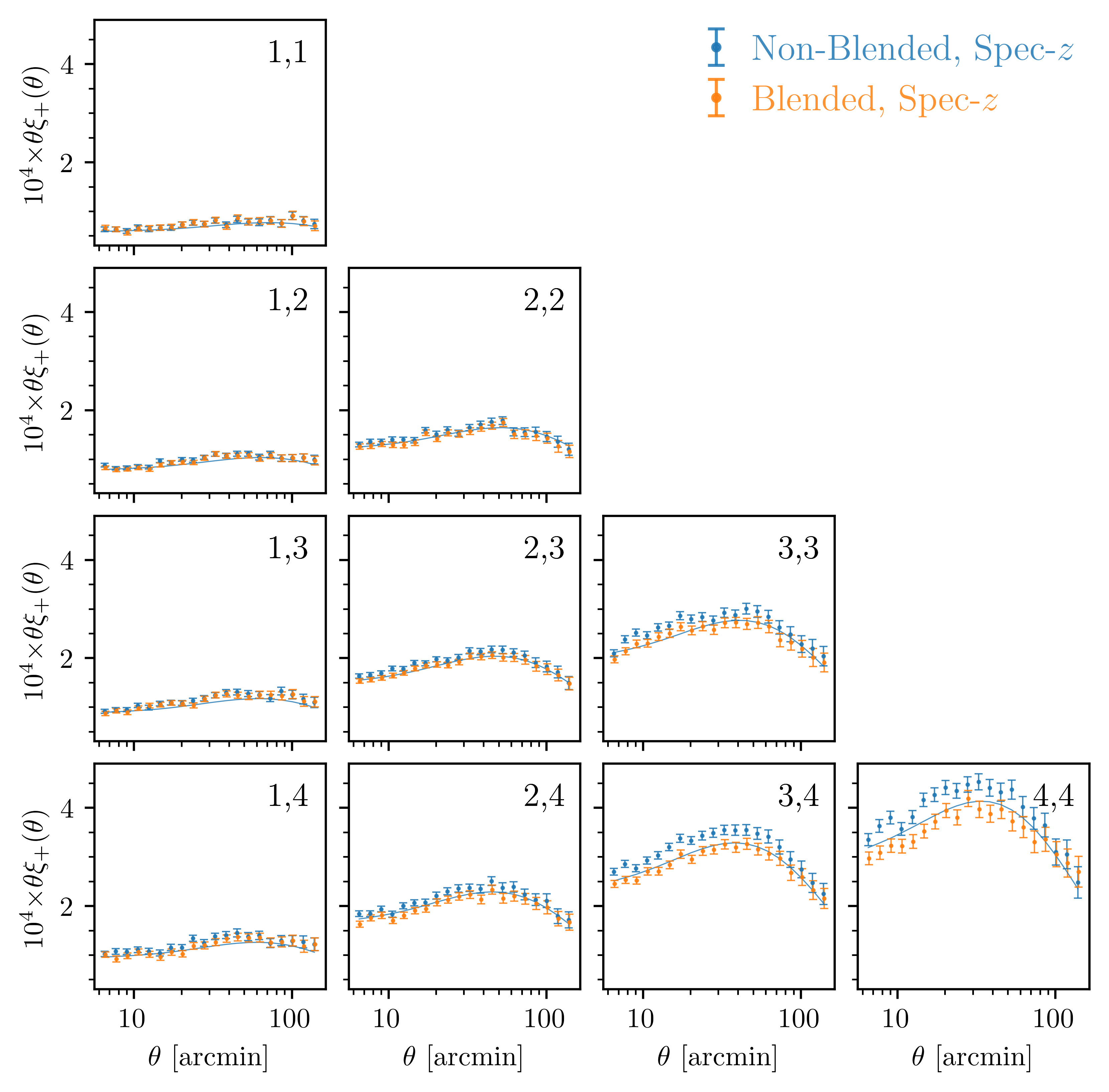}
        \caption{}
        \label{zsnbvszsb1} 
    \end{subfigure}

	\begin{subfigure}[b]{0.59\textwidth}
        \includegraphics[width=1\textwidth]{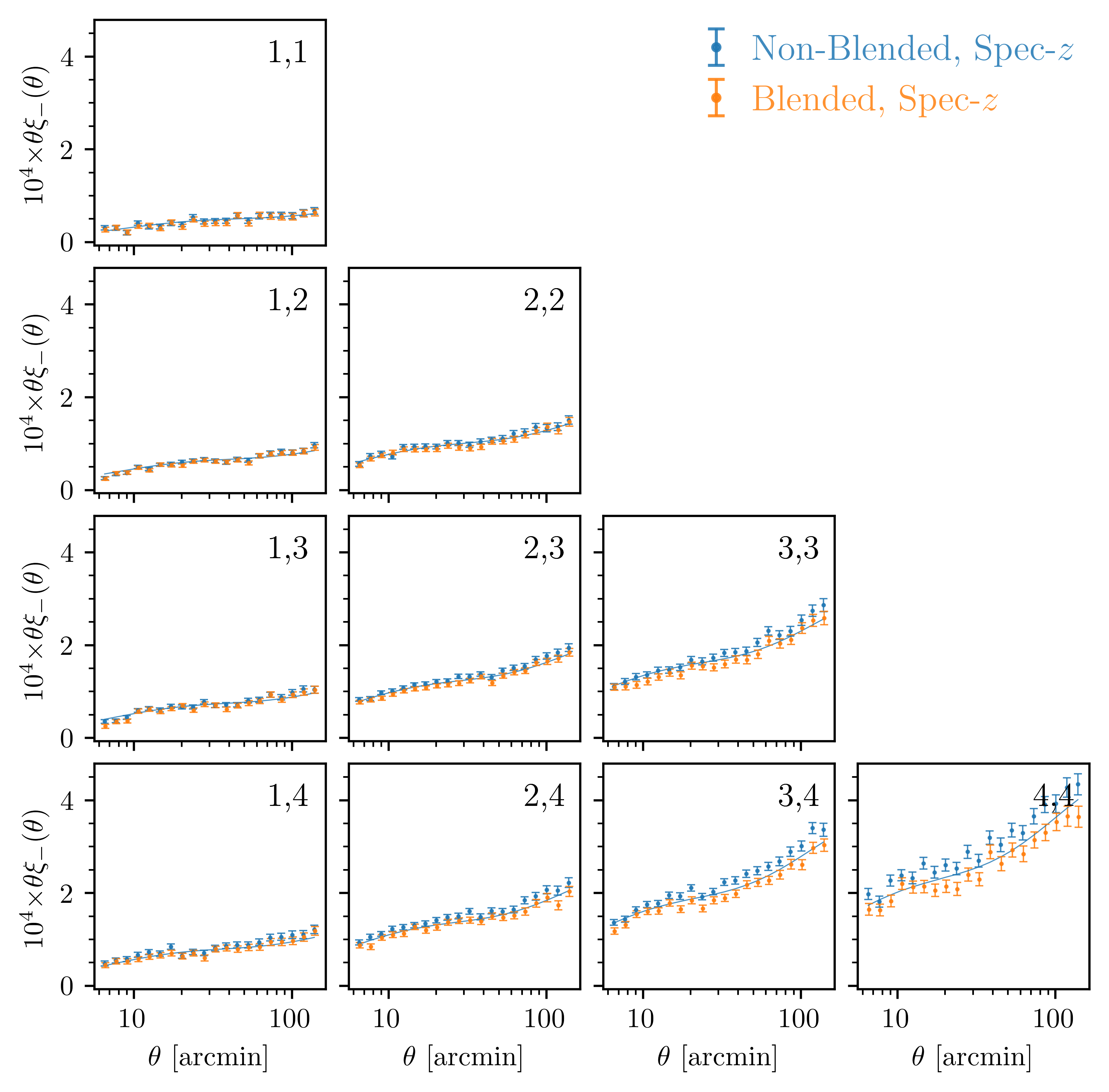}
        \caption{}
        \label{zsnbvszsb2} 
    \end{subfigure}
    
	\caption{Shear--shear cross-correlation (a) $\xi_+$ and (b) $\xi_-$ between spec-$z$ bins measured in \texttt{Buzzard}, for the no-blended and blended source samples. The diagonal panels show the autocorrelation. Error bars are obtained from 2\,000 jackknife resamples. Overlaid are the fiducial theory lines (not fitted to data) with no shear calibration parameter applied where we use the redshift distribution of the non-blended source sample. A suppression of correlation due to the effects of blending is noticeable.}
	\label{zsnbvszsb}
\end{figure*}

\begin{figure*}
	\centering
	\begin{subfigure}[a]{0.59\textwidth}
         \includegraphics[width=1\linewidth]{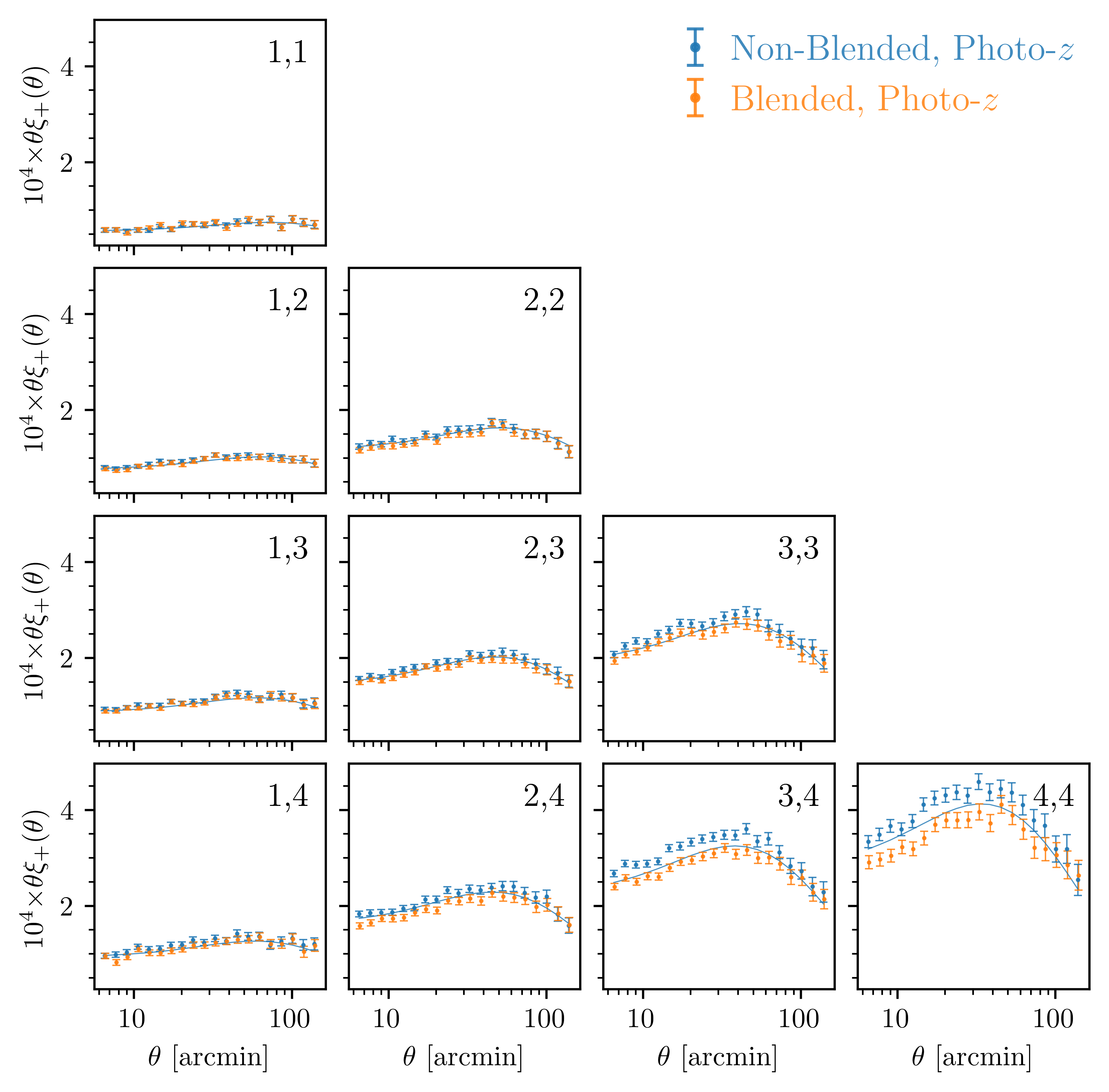}
         \caption{}
         \label{zpnbvszpb1} 
    \end{subfigure}

	\begin{subfigure}[b]{0.59\textwidth}
         \includegraphics[width=1\textwidth]{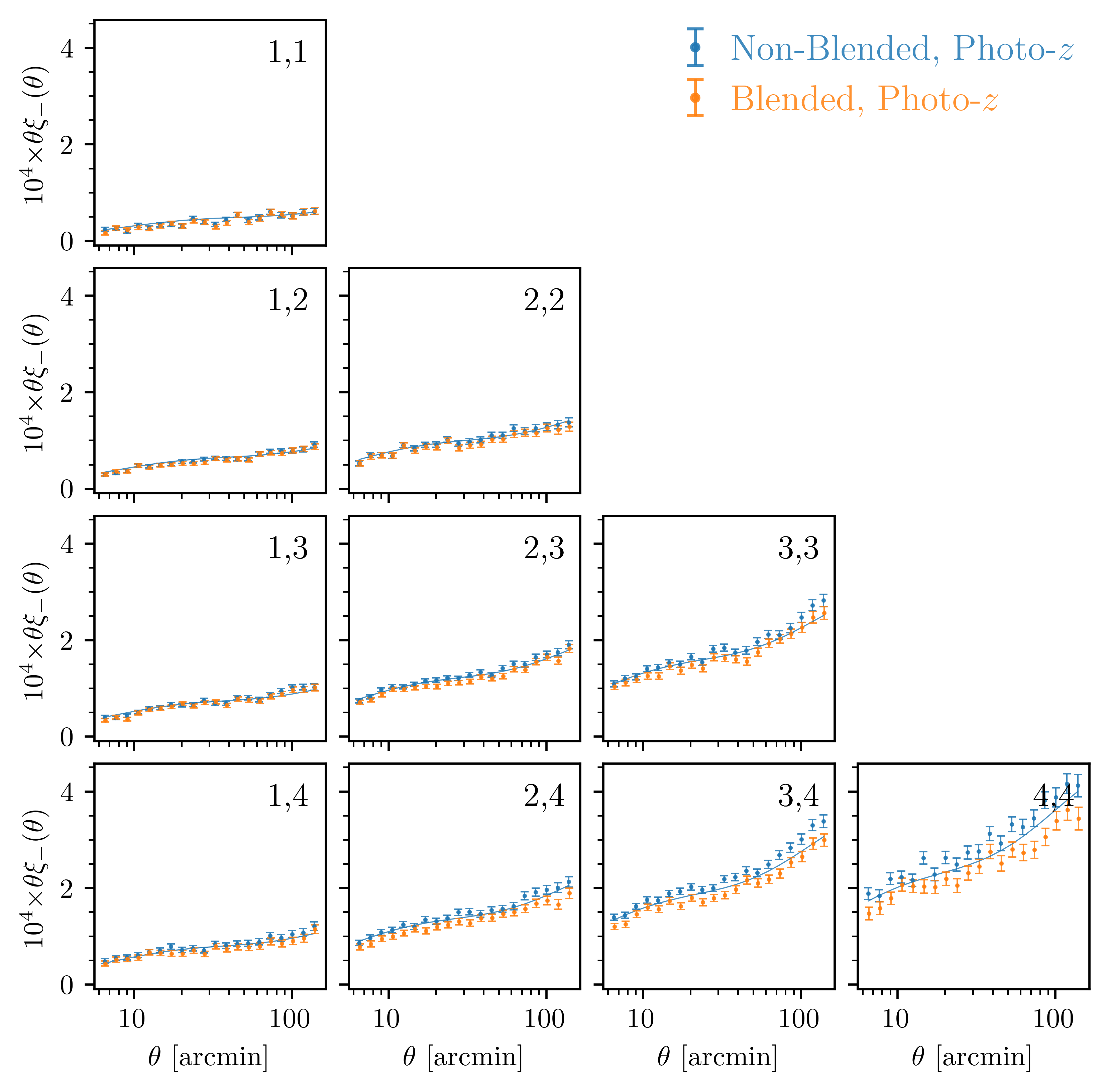}
         \caption{}
         \label{zpnbvszpb2}
    \end{subfigure}
    
	\caption{Shear--shear cross-correlation (a) $\xi_+$ and (b) $\xi_-$ between photo-$z$ bins measured in \texttt{Buzzard}, for the no-blended and blended source samples. The diagonal panels show autocorrelation. Error bars are obtained from 2\,000 jackknife resamplings. Overlaid are the fiducial theory lines (not fitted to data) with no photo-$z$ bias or shear calibration parameter applied where we use the redshift distribution of the non-blended source sample.  As in Fig. \ref{zsnbvszsb}, a suppression of correlation due to the effects of blending is noticeable.}
	\label{zpnbvszpb}
\end{figure*}

To quantify the cosmic shear information using two-point statistics, we use equation \eqref{eq:cosmicshear} to compute the shear--shear correlation between the shapes of source galaxies after analytically deconvolving from them the $i$-band PSF in the moment space by solving equation \ref{qconv} for $Q^{\prime}_{ij}$ -- this time with noisy galaxy moments -- while assuming the flux is preserved. Recall that the PSF is circular and Gaussian with FWHM $\sim0.9~ \rm arcsec$, in keeping with what we had assumed for the convolution process. The cosmic shear measurement is done across spectroscopic and photometric redshift bins, both averaged over two \texttt{Buzzard} realizations.

In order to account for the uncertainty in the ellipticity measurement, we weight our correlations according to the shape noise per tomographic bin\footnote{The variation in shape noise for the different tomographic bins is not large for our source samples. For the non-blended source sample with spec-$z$ tomography the mean shape noise is on average 0.21 per component with a standard deviation of 0.002 among four bins.} ($\sigma_{\rm tomo}$) and shape error ($\delta e$) of our galaxies. The galaxy weight for each galaxy is defined as the inverse of the sum of the squares of the shape noise and shape error (values are per-component):
\begin{equation}
    w_i = \frac{1}{\sigma_{\rm tomo}^2+(\delta e_i)^2},
    \label{wi}    
\end{equation}
where the subscript $i$ refers to the $i$th galaxy.
We exclude from our analysis galaxies with the deconvolved ellipticity $e>0.8$, $\delta e >0.25$, $i$-band flux $\rm S/N < 10$ and the trace of the second moments' matrix ($T=Q_{11}+Q_{22}$) of the observed shape (not deconvolved) less than 1.5 times that of the PSF. The error in shape measurement ($\delta e$) is estimated as the absolute magnitude of the difference between the observed (noisy) ellipticity and the noise-free (lensed only) ellipticity. Our observed shapes are calculated in terms of the $i$-band $\rm S/N$ (Section \ref{galprop}), so are the associated errors ($\delta e$'s).

In addition, we restrict ourselves to galaxies detected in at least four LSST bands, including the $i$ band with $i>22$ drawn from our cosmology sample already with the $i<24$ magnitude cut. We have thus come to the \texttt{Buzzard} source sample, which we will make separately for different cases (depending on blending and tomography) and use for cosmological inference. Using these quality cuts ensures reliable shape measurements and photometry while maintaining a high shear signal for the source sample. Including brighter galaxies ($i<22$) would only add $\sim$28 per cent more galaxies to the cosmology sample and $\sim$42 per cent more galaxies to the source sample given the steep apparent magnitude vs number counts relation. We are aiming at the background galaxies and therefore it is more appropriate to select fainter galaxies for that purpose. A recent example of such a cut at the $22{\rm nd}$ magnitude is \cite{ihasan2022}.
The fraction of galaxies in the \texttt{Buzzard} cosmology sample that satisfy all the criteria (i.e., will be kept as the source sample) is about 44 per cent for both the non-blended and blended cases.

We use \texttt{TreeCorr}\footnote{\url{https://github.com/rmjarvis/TreeCorr}} \citep{shearm2} to compute the correlation functions in the source sample for all redshift bin combinations (made from four redshift bins shown in Fig. \ref{tomobins}) in $N_{\rm th}=20$ angular bins, logarithmically spaced between 6 and 150 arcmin while setting \texttt{bin\_slop}$~= 0$ for a high level of accuracy identical to brute-force calculation\footnote{Acoording to \texttt{TreeCorr}'s documentation, \texttt{bin\_slop} specifies the maximum possible error any pair of points can have, given as a fraction of the bin size. For more information see \url{https://rmjarvis.github.io/TreeCorr/_build/html/binning.html}.}. Correlations are averaged over two realizations. Figs \ref{zsnbvszsb} and \ref{zpnbvszpb} show the calculated shear--shear correlations, $\xi_{+}(\theta)$ and $\xi_{-}(\theta)$, for several spec-$z$ and photo-$z$ tomographic bin combinations in our analysis with and without the effects of blending. Error bars are obtained from 2\,000 jackknife resamplings (1,000 per mock realization) described in Section \ref{covariance}. The theory prediction lines (not fitted to data) from our fiducial $\Lambda$CDM model are produced using the Core Cosmology Library\footnote{\url{https://github.com/LSSTDESC/CCL}} (\texttt{CCL}; \citealt{ccl}). One expects non-zero correlations in off-diagonal panels due to foreground large-scale structure dark matter lensing galaxies across redshift bins. Overall, the correlations suggest that there is suppression from blending across all panels.

Galaxy--galaxy blending systematically changes several different types of measurements over a large volume -- data of dimension much higher than the one of blending. There are $2 \times \frac{N_{\rm tomo}(N_{\rm tomo}+1)}{2}$ independent two-point shear--shear correlation functions, defined in $N_{\rm tomo}$ redshift bins (the factor of 2 accounts for the fact that shear is a spin-2 field requiring two kinds of correlations to describe it in real space, i.e. $\xi_+$ and $\xi_-$). Given $N_{\rm tomo}=4$ and $N_{\rm th}=20$, the total number of data points for our cosmic shear data vector, $\bar{v}$, comes to $N_{\bar{v}}=400$.
Although there is information in the combined analysis of these many correlations vs redshift, it may still be insufficient to fully suppress the noise arising from unrecognized blending of galaxies that propagates to cosmology. We investigate this in Section \ref{cosmoparam} by analysing systematic changes in cosmological inference considering blending effects in the cosmological volume of the \texttt{Buzzard} simulations.

\section{Covariance}\label{covariance}
In cosmological parameter estimation using the weak lensing probe, it is crucial to construct a reliable covariance matrix that includes the effects of both systematic errors and sample variance. 
\cite{taylor2013} show that the accuracy of parameter estimation is limited by the accuracy of the inverse data covariance matrix -- the precision matrix.
In addition to calculating the correlation functions, a hybrid of jackknife+mock method\footnote{The delete-one jackknife pseudo-samples are supposed to be resulting from the entire set minus one patch at a time. Since multiple mock realizations are involved, the approach we present here does not exactly replicate what jackknife does, but it is closer to the expected answer than simply averaging individual jackknife covariance matrices for each realization (e.g. \citealt{Escoffier2016}).} is employed to internally estimate the covariance between the correlations obtained from $N_{\rm r}$ mock realizations. This is achieved by replacing every $i \rm{th}$  jackknife pseudo-data vector (calculated from each jackknife pseudo-sample) from the $j \rm{th}$ realization, $v_i^{j}$, by the weighted average of $v_i^{j}$ and the full data vectors from $N_{\rm r}-1$ other realizations, that is
\begin{equation}
    \label{jkmock}
    \mathcal{V}_{i}^{j} = \frac{(N_{\rm jk}^j-1) v_i^{j} + \sum_{\substack{k=1, k \neq j}}^{N_{\rm r}} N_{\rm jk}^k \bar{v}^k }{ (N_{\rm jk}^j-1)+\sum_{\substack{k=1, k \neq j}}^{N_{\rm r}} N_{\rm jk}^k }, ~ i \in \{1,...,N_{\rm jk}^j\}
\end{equation}
and then calculating the delete-one jackknife covariance matrix using a reconstructed ``design matrix'' made by concatenating these weighted pseudo-data vectors taken from all realizations as
\begin{equation}
    \mathcal{V} = 
    \begin{bmatrix}
        \mathcal{V}_1^1 \\
        \vdots \\
        \mathcal{V}_{N_{\rm jk}^1}^1 \\
        \vdots \\ 
        \mathcal{V}_{\rm 1}^{N_{\rm r}} \\ 
        \vdots \\ 
        \mathcal{V}_{N_{\rm jk}^{N_{\rm r}}}^{N_{\rm r}} 
    \end{bmatrix}.
\end{equation}
Here $N_{\rm jk}^j$ is the number of patches used for jackknifing in the $j \rm{th}$ realization and $\bar{v}^k$ is the full data vector of the $k \rm{th}$ realization.
With the same number of jackknife patches, $N_{\rm jk}$, used for each realization, equation \eqref{jkmock} can be simplified as
\begin{equation}
    \label{jkmock2}
    \mathcal{V}_{i}^{j} = \frac{N_{\rm jk} \big( v_i^{j} + \sum_{\substack{k=1, k \neq j}}^{N_{\rm r}} \bar{v}^k \big) - v_i^{j}  }{ N_{\rm r} N_{\rm jk}-1 }, ~ i \in \{1,...,N_{\rm jk}\}.
\end{equation}

The size of jackknife patches has traditionally been restricted to larger than the scale of the two-point correlation function of concern. Nevertheless, according to a recent study by \cite{favole2021}, when using jackknife as the covariance estimator for cosmological analyses, it is more beneficial to beat down the noise by building many resamplings than to select a jackknife size greater than e.g. the maximum galaxy clustering scale. We chose 1000 equal-area jackknife patches over 10\,254 square degrees in our source sample using the \texttt{kmeans} algorithm via \texttt{TreeCorr} that leads to an effective jackknife scale of 3.2 deg. Hence, there is an adequate number of jackknife patches while their effective diameter (3.2 deg) is greater than the maximum angular scale of our correlation measurements (2.5 deg).
In this work, $N_{\rm r}=2$ and $N_{\rm jk}=1,000$, so the number of jackknife replicates totals $n^{\prime}=N_{\rm r} N_{\rm jk}=2000$ for two realizations. As it turns out, this is desirably five times larger than the length of the data vector, 400, which makes it less likely for the resulting covariance matrix to be unreasonably noisy or singular. The 3D design matrix, $\mathcal{V}$, is then fed to \texttt{TreeCorr} functions to generate the 2D jackknife covariance matrix, whose $i$th row and $j$th column reads
\begin{equation}
C_{ij} = \frac{ n^{\prime} - 1 }{ n^{\prime} } \sum_{k=1}^{ N_{\rm r} } \sum_{l=1}^{ N_{\rm jk} } (\mathcal{V}_{l, i}^{k} - \bar{\mathcal{V}}_i) (\mathcal{V}_{l, j}^{k} - \bar{\mathcal{V}}_j),
\end{equation}
where $i$ and $j$ on the right-hande side index the data points in the pseudo-data vectors, and $\bar{\mathcal{V}}_i$ is the $i$th correlation data point averaged over all weighted jackknife pseudo-data vectors from all realizations. Fig. \ref{mysample} shows the jackknife regions for one realization of our source sample.

\begin{figure}
	\centering
   \includegraphics[width=\linewidth]{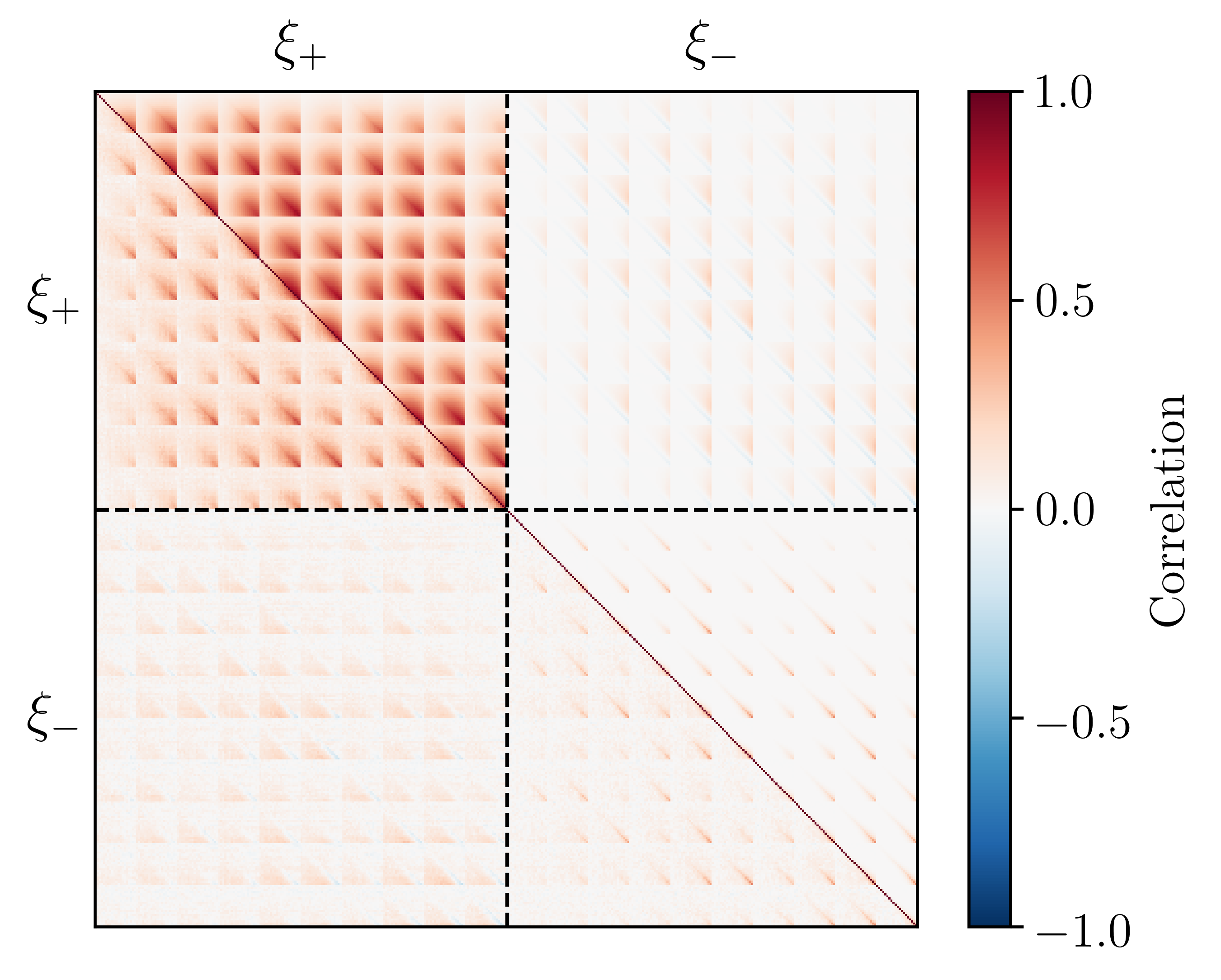}
	\caption{Cosmic shear correlation matrix (normalized covariance matrix, $C_{ij}/\sqrt{C_{ii}C_{jj}}$, where $C_{ij}$ are the elements of the covariance matrix) for the non-blended case with spec-$z$ tomography produced by our jackknife estimator ({\it lower left} triangle) and the analytical Gaussian model ({\it upper right} triangle). We use the former in this study with a total of 2000 jackknife pseudo-samples from two realizations of \texttt{Buzzard} sources (1000 pseudo-samples per realization). See equation \eqref{jkmock} and its description. In Appendix \ref{cosmosis_gcov}, we repeat the analysis with the Gaussian covariance matrices produced separately for each sample and compare the results.}
	\label{jkcov} 
\end{figure}

The covariance matrices used in cosmic shear studies are typically large since they incorporate many tomographic bin combinations and many angular bins per combination. For the internal covariance estimators such as jackknife, this requires a sample of billions of galaxies to maintain a high S/N. With this in mind, we will also use an {\it analytical} approach to construct a Gaussian covariance matrix using the \texttt{TJPCov}\footnote{\url{https://github.com/LSSTDESC/TJPCov}} code for comparison. It utilizes the techniques outlined in \cite{singh21} and \texttt{Skylens}\footnote{\url{https://github.com/sukhdeep2/Skylens\_public}}, and ignores the non-Gaussian features whose effects on final parameter constraints are very small.
Since the evaluation of the Legendre polynomials is extremely slow at low angles (high $\ell$'s), we use a flat-sky approximation that transforms them into Bessel functions.
An example of the covariance matrices used in this study is shown in Fig. \ref{jkcov}. The jackknife and the Gaussian covariance matrices are shown, respectively, in the lower and upper triangles of the same plot for comparison. Here, we demonstrate that our internally estimated covariance matrix is comparable, at least pattern-wise, with the analytical covariance matrix.

\section{Blending and Cosmological Inference}\label{cosmoparam}

\subsection{Analysis setup}
The goal of this study is to perform a cosmic shear analysis using all the correlation functions across the redshift bins to evaluate the impact of blending on cosmological parameter estimation. A joint analysis of all the LSST probes is expected to increase the precision \citep{zhantyson17},  though this analysis is post-poned to a later study.
We find the likelihoods comparing the theoretical and experimental correlation functions using the Markov chain Monte Carlo (MCMC) technique for which we provide a covariance matrix from the jackknife resampling on the data and the analytical covariance matrix explained in the previous section.

This analysis is carried out using \texttt{CosmoSIS}\footnote{\url{https://bitbucket.org/joezuntz/cosmosis}}, a modular pipeline developed by \citet{zuntz} that performs the Bayesian inference and provides constraints on the cosmological parameters. In order to generate theory predictions, the matter power spectra are computed using the \texttt{CAMB}\footnote{\url{http://camb.info}, January 2015 version} \citep{camb2000, camb2012} Boltzmann module with the \texttt{HALOFIT} \citep{halofit2003, halofit2012} non-linear correction. We employ the \texttt{MultiNest} sampler \citep{feroz09} with 500 \texttt{livepoints}, \texttt{efficiency} set to 0.1, \texttt{tolerance} set to 0.1 and \texttt{constant efficiency} set to \texttt{False}. The reader is referred to \cite{feroz09} for a detailed description of these parameters. The contours and confidence intervals in the parameter space are then extracted and plotted using \texttt{ChainConsumer} \citep{chainconsumer} for both the jackknife covariance matrix and the Gaussian covariance matrix. The contour plots for the jackknife covariance matrix are displayed in the main text while the ones for the Gaussian covariance matrix can be found in Appendix \ref{cosmosis_gcov}.

The priors used for parameter constraints are shown in Table \ref{priors}.  This table consists of non-informative priors for cosmological parameters of interest to ensure that they will not bias our inferences. However, for the systematic parameters associated with measurements of shear (the shear calibration bias, $m_{\mathrm{1-4}}$) and photometric redshift (the photo-$z$ shift bias, $\Delta z^{\mathrm{1-4}}$), we use our \textit{a priori} knowledge to impose informative priors. For the shear calibration parameter, we take the ensemble average of the observed (sheared, convolved, noisy, and then deconvolved) ellipticity $e_{\rm obs}$ and sheared-only ellipticity $e_{\rm true}$ of the \texttt{Buzzard} source sample:
\begin{equation}
    \langle e_{\rm obs} \rangle = (1+m) \langle e_{\rm true} \rangle.
\end{equation}
This results in a value for $m$ of $\sim0.02$. It is not required to maintain this value across all tomographic bins and for various samples since it depends on multiple factors including galaxy flux, redshift, and size distribution, and is statistically correlated between tomographic bins. We, therefore, assign a conservatively wide Gaussian prior of $\sigma_{\rm m}=0.03$ for the shear calibration parameter in all tomographic bins. In effect, for each pair of tomographic bins, the theoretical predictions of shear--shear correlations are re-scaled as
\begin{equation}
    \xi^{ij}_\pm(\theta) \rightarrow (1+m_{i})(1+m_{j}) \, \xi^{ij}_\pm(\theta),
\end{equation}
where $i$ and $j$ denote different tomographic bins. For the photo-$z$ shift bias, we cross-correlate our observed source sample with the validation sample of $\sim200,000$ galaxies with corresponding spec-$z$ values to determine the bias in the mean of the redshift distribution, $\Delta z$, along with its standard deviation derived as

\begin{equation}
    S_{\Delta z} = \sqrt{ S^2_{\rm s}+S^2_{\rm p}-{\rm Cov}(s,p)}\,,
\end{equation}
where $S_{\rm s}=\sigma_{\rm s}/n$ and $S_{\rm p}=\sigma_{\rm p}/n$ with $\sigma_{\rm s}$ and $\sigma_{\rm p}$ being the weighted standard deviations of the spectroscopic and photometric samples, respectively (see equation \ref{wi} for the weighting scheme).
Here $n$ is the number of galaxies cross-correlated and ${\rm Cov}(s,p)=2rS_{\rm s} S_{\rm p}$ with $r$ being the Pearson correlation coefficient of spec-$z$ and photo-$z$ point estimates. This is necessary because the samples are correlated. This calculation is conducted for each tomographic bin resulting in $\Delta z^i$ values where $i \in \{1,2,3,4\}$ as seen in Table \ref{priors}. The stacked photo-$z$ PDFs of galaxies ($p(z)$ in Fig. \ref{dndz_s_photoz_pdf}) are fed to \texttt{CosmoSIS}, in place of the redshift number density, $n(z)$, for the cases with photo-$z$ tomography. The uncertainties in the redshift distribution mean values (weighted according to equation \ref{wi}) transform the redshift distribution of the $i$th tomographic photo-$z$ bin as
\begin{equation}
    n^{i}(z) \rightarrow n^{i}(z - \Delta z^{i}).
\end{equation}

It should be noted that all priors are adjusted based on the non-blended source sample with spec-$z$ tomography (we will not change them at any point in our analysis) to serve as a reference for how the inference changes due to blending and with the use of photometric redshifts for tomography.

We generate runs with and without blending using all the data points available in our two-point correlation functions. Since baryonic effects are not present in \texttt{Buzzard}, no additional angular scale cuts are imposed on our data vectors to remove these effects. Moreover, the lower bound of our angular scale (6 arcmin) is large enough to minimize the impact of finite simulation resolution on the cosmic shear signal discussed in \cite{derose}. Our results may be optimistic in terms of constraining power, as baryonic effects will always be present in the real universe, but this is less relevant to the primary purpose of this study, which is to compare cosmological findings of blended vs. non-blended universe.

In the course of our analysis, we look only at constraints on parameters most crucial to the cosmic shear signal: the mean matter density in the universe, $\Omega_{\rm m}$, the amplitude of matter density fluctuations on scales of 8 $h^{-1} \rm Mpc$, $\sigma_8$, and a combination of these two parameters defined as $S_8=\sigma_8 (\Omega_{\rm m}/0.3)^{0.5}$ to break the degeneracy between $\Omega_{\rm m}$ and $\sigma_8$. 

\subsection{Unrecognized blending bias in cosmological inference}
\begin{figure}
	\centering
	\includegraphics[width=\linewidth]{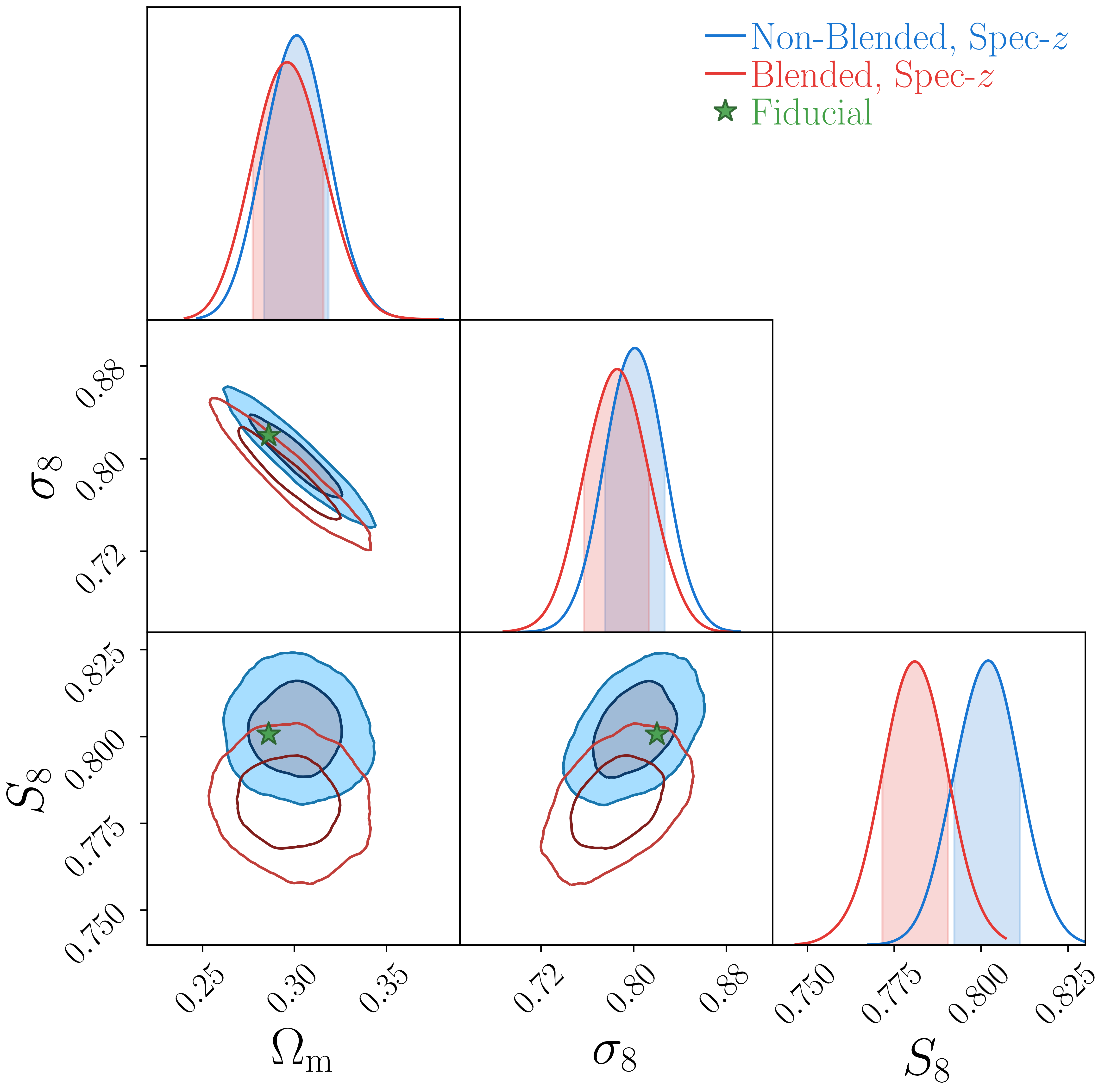}
	\caption{Cosmological parameters estimated using cosmic shear for the non-blended vs blended source sample for tomographic bins selected using spectroscopic redshifts. The covariance matrix is produced by the jackknife estimator separately for each sample. The contours of the blended and non-blended cases are displayed in filled blue and unfilled red, respectively. In each case, the contours drawn correspond to $1\sigma$ (68 per cent) and $2\sigma$ (95 per cent) confidence levels. The green star marks the fiducial parameters.  The inclusion of blending causes a noticeable shift in our parameter estimates, particularly in $S_8$. This bias has an additional effect baked into it from the impact of blending on the spec-$z$'s themselves (our choice of the spec-$z$ of the brightest galaxy noted in Section \ref{photozest}) that compounds the amplitude of the shift.}
	\label{cosmoz}
\end{figure}

\begin{figure}
	\centering
	\includegraphics[width=\linewidth]{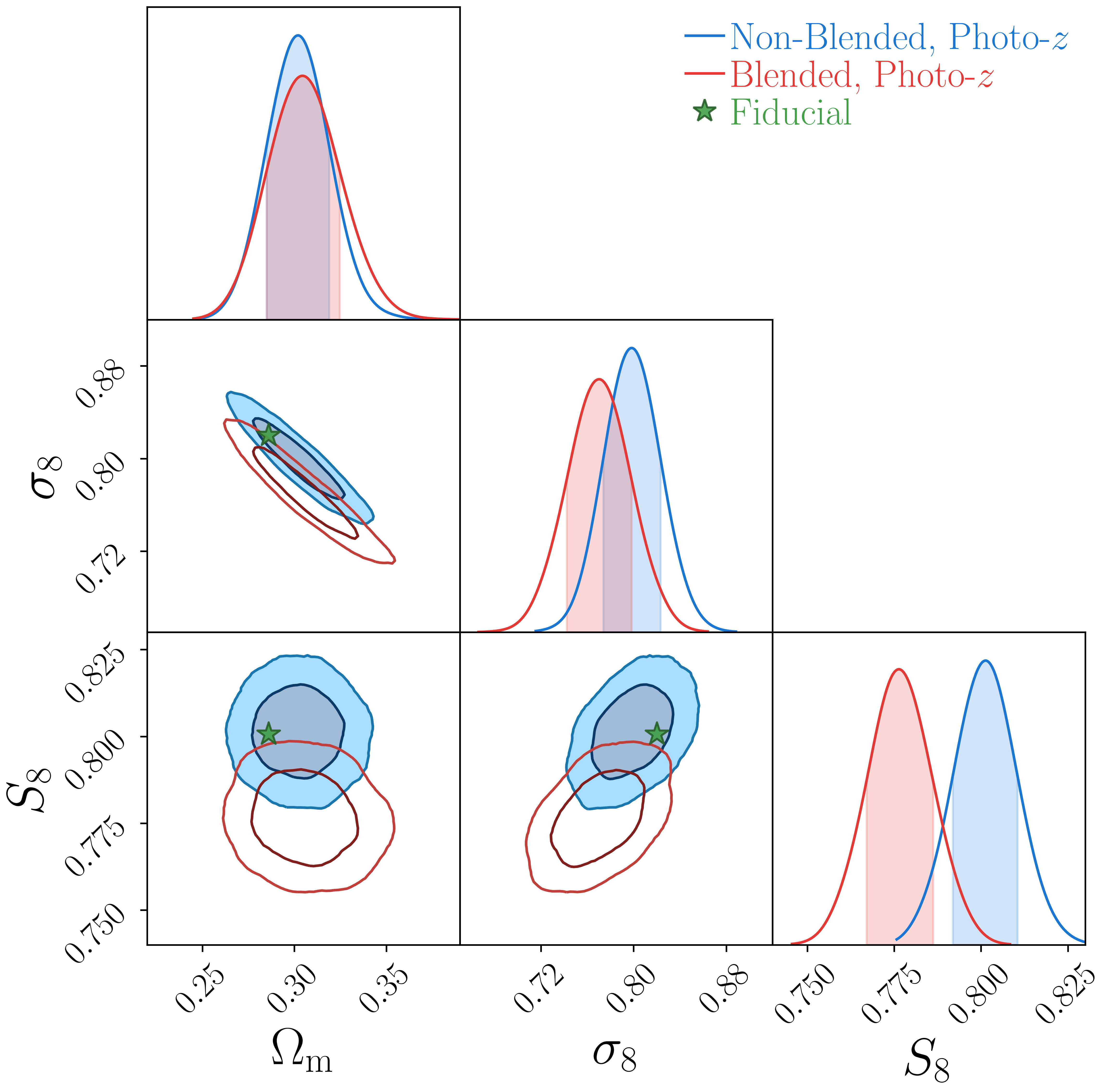}
	\caption{Same as Fig. \ref{cosmoz} but for tomographic bins selected using photometric redshifts. We observe a parameter estimate shift similar to that seen in Fig.~\ref{cosmoz}. Again, added to this bias is the impact of blending on the photo-$z$'s themselves (because of the use of collective photometric data of the blend for estimating redshifts) that compounds the amplitude of the shift. This bias is most visible in the bottom right panel where we show $S_8$ posteriors.}
	\label{cosmopz}
\end{figure}

The estimated cosmological parameters for both the non-blended and blended source samples are shown in Figs \ref{cosmoz} and \ref{cosmopz} using spec-$z$ and photo-$z$ tomography, respectively, where we employ jackknife covariance matrices, and probe cosmology only with shear--shear correlations. Clear bias is visible in the $S_8\mathrm{-}\Omega_{\rm m}$ plane. While blending affects the covariance of cosmological parameters, in both the spec-$z$ tomography and photo-$z$ tomography the difference between the blended case and non-blended case can be less than (in the cases of $sigma_8$ and $Omega_{\rm m}$) or greater than (in the case of $S_8$) the sum of the corresponding $1\sigma$ posterior uncertainties. As far as statistics are concerned, this is still a relatively small change, and gives some hope that one could model and correct for the effects of blending for unrecognized blends in a well-defined sample. 

\begin{table*}
    \centering
    \makebox[\linewidth]{
    \resizebox{1.01\linewidth}{!}{%
    \begin{tabular}{l  l  l  l  l  l }
        \toprule
        {Tomographic} & {Redshift range}  & {\# of source galaxies in the} & {\# of source galaxies in the} & {\# of source galaxies in the} & {\# of source galaxies in the} \\ 
        {bin number} & & {spec-$z$ sample (non-blended)} & {photo-$z$ sample (non-blended)} & {spec-$z$ sample (blended)} & {photo-$z$ sample (blended)} \\
        \midrule
        1 & 0.3 -- 0.5 & 73,585,956.5  & 85,873,670.5 & 69,800,775 & 80,357,357.5  \\
        2 & 0.5 -- 0.7 & 76,273,201  & 70,510,758 & 73,946,380 & 69,636,192 \\
        3 & 0.7 -- 0.9 & 55,982,626  & 60,930,782.5 & 55,945,993.5 & 60,773,952 \\
        4 & 0.9 -- 1.2 & 30,693,222.5  & 29,936,703 & 32,550,779.5 & 32,294,214.5 \\
        \midrule
        {Total} & {0.3 -- 1.2} & {236,535,006}  & {247,251,914} & {232,243,928} & {243,061,716} \\
        \bottomrule
    \end{tabular}
    }%
    }
    \caption{Table of the number of galaxies {\it per realization} of the \texttt{Buzzard} simulation after weak lensing sample cuts (described in Section \ref{cosmicshearprobe}) for different tomographic redshift bins. We use two total realizations in this study. Tomographic binning is based on both the spectroscopic and photometric redshift. These galaxies are used to compute the correlation functions for our cosmic shear analysis.}
    \label{source_table}
\end{table*}

Our attention is drawn to the fact that the number of galaxies lost due to blending is partially offset by the unrecognized blended objects that collectively exceed the magnitude limit, i.e. by becoming brighter than our upper magnitude cut at $i=24$ for the source sample. Table \ref{source_table} tabulates the final number counts after applying all the quality cuts mentioned in Section \ref{cosmicshearprobe}. This shows an overall decrease in the number of galaxies due to blending regardless of the method used to assign galaxies to tomographic redshift bins. While a decrease in $\sigma_8$ is seen due to blending in Figs \ref{cosmoz} and \ref{cosmopz}, $\Omega_{\rm m}$ is not as affected by it, and neither is by the loss of galaxies in the sample.
By emulating blends in the catalogue we are smoothing out shapes, i.e. shears in the statistical sense, over the ensemble. In other words, blending introduces altered (collective) shapes for the unrecognized blends in ways that are unrelated to lensing.  This tends to dilute the lensing signal because the random nature of the blends does not have a preferred direction. Consequently, this lowers the amplitude of the correlations between the shapes meaning the amplitude of matter fluctuations, $\sigma_8$, is lower when blending is present, and so is the normalized version of it, $S_8$.
The analysis above is repeated with analytical Gaussian covariance matrices and shows similar biases. For more details see Figs \ref{cosmoz_gausscov} and \ref{cosmopz_gausscov}.

\subsection{Effect of omitting recognized blends}\label{recbl}
Since the separation angle threshold ($\theta_{\rm b}=4~ \rm arcsec$) is larger than the sum of the half-light radii of an average galaxy pair in the \texttt{Buzzard} cosmology sample ($\sim1.55~ \rm arcsec$), the surface brightness shows a detectable dip between the centroid positions of some overlapping galaxies in the blended scenes. Therefore, blended galaxies in these instances are recognized blends where detection algorithms are able to separate objects that have been blended together and attempt to deblend them \citep{scarlet}. 
In very deep imaging such as planned for LSST, the low surface brightness tails of galaxies may extend to the mean galaxy--galaxy separation on the sky out to several arc seconds. \citet{bosch2018} find that nearly 60 per cent of galaxies in HSC Wide survey images are extracted from blended groups in above-threshold regions of the sky. In general, deblending closely spaced galaxies can be trusted more for pure photometry than for shape measurement. It can cause systematic biases in determining galaxy shapes along the direction connecting the (recognizably) blended galaxies. Therefore, at least in the case of shear-based precision cosmology, it may be prudent to eliminate severe cases of recognized blends from the source sample.

We define the purity\footnote{Purity relates to ``blendedness'' as defined in the LSST Data Management (DM) software stack \citep{lsstdmsys}, by the relation $\rho=1-$blendedness.} $\rho$ of a galaxy as a measure of the severity of its blending with overlapping galaxies as \citep{sanchez21},

\begin{equation}
\rho_i \equiv \frac{\sum_p s_{ip}\cdot s_{ip}}{\sum_p (s_{ip}\cdot \sum_j s_{jp})} \; ,
\label{eqn:rho}
\end{equation}
where $p$ sums over all pixels that belong to the galaxy of interest $i$ above a limiting surface brightness where the signal is at least 5 per cent of the sky noise, and $j$ sums over all galaxies that overlap $i$, including $i$ itself. 
In other words, purity is a ratio of weighted pixel sums and includes a range $0 < \rho_{i} \leq 1$ with $\rho_{i} = 0$ for completely blended and $\rho_{i} = 1$ for completely isolated galaxies. This quantity tells us how much of a galaxy's surface brightness profile overlaps with other sources' surface brightness profiles weighted by flux. In this respect, it is not symmetric between overlapping pairs.

As a proxy for identifying catastrophic recognized blends, we assume a minimum purity threshold for galaxies that we keep in the source sample, i.e. we discard the less pure ones.  In practice, the purity threshold is set by the limiting surface brightness of the survey and the method used to detect and deblend the sources. If we flag galaxies in crowded regions with $\rho<0.95$ as catastrophic recognized blends, we retain $\sim 95$ per cent of the total cosmology sample, or $\sim 94$ per cent of the total source sample. Note that galaxies that are not in the crowded regions (for which we made image cutouts) do not have any neighbours within $4~ \rm arcsec$. Therefore, it is safe to assume that they all have the purity of 1, thereby being isolated according to the adopted scheme.

We shall assume that the majority of recognized blends are properly deblended and do not need to be thrown out of the sample. For typical galaxies in our \texttt{Buzzard} source sample, the pre-convolution half-light radii range between 0.5 and 1 arcsec, and convolved with our $\sim 0.9 ~ \rm arcsec$ effective PSF FWHM, pairs of such galaxies are likely to show large systematic distortions in their second intensity moments for pair distances out to a few arcseconds.
However, to reiterate what we pointed out in the previous paragraph, at larger separation than our adopted threshold for the blended scenes ($\theta_{\rm b}=4~ \rm arcsec$), a typical pair of galaxies may be deblended accurately for purposes of shape measurement at the sub per cent shear systematic level.

\begin{figure}
	\centering
    \includegraphics[width=\linewidth]{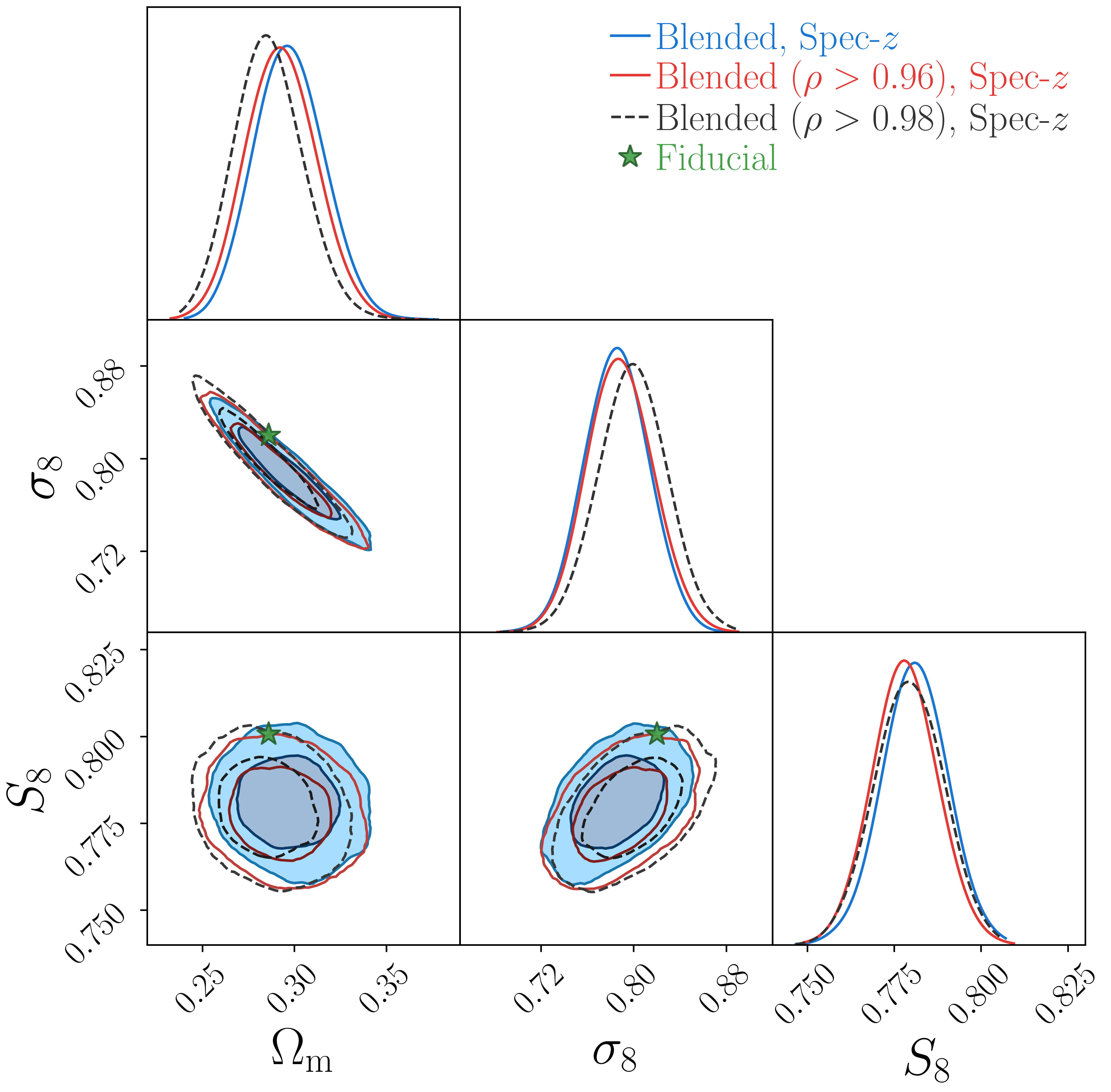}
	\caption{Cosmological parameter estimation using cosmic shear with spec-$z$ tomography, before and after the exclusion of recognized blends from the source sample based on their purity values ($\rho$ in equation \ref{eqn:rho}) in the blended scenes. The covariance matrix is produced by the jackknife estimator separately for each case. The contours of the blended, blended with $\rho>0.96$ cut, and blended with $\rho>0.98$ cut samples are displayed in filled blue, unfilled red (solid line) and unfilled black (dashed line), respectively. In each case, the contours drawn correspond to $1\sigma$ (68 per cent) and $2\sigma$ (95 per cent) confidence levels. The green star marks the fiducial parameters.}
	\label{cosmoz_rb} 
\end{figure}

\begin{figure}
	\centering
    \includegraphics[width=\linewidth]{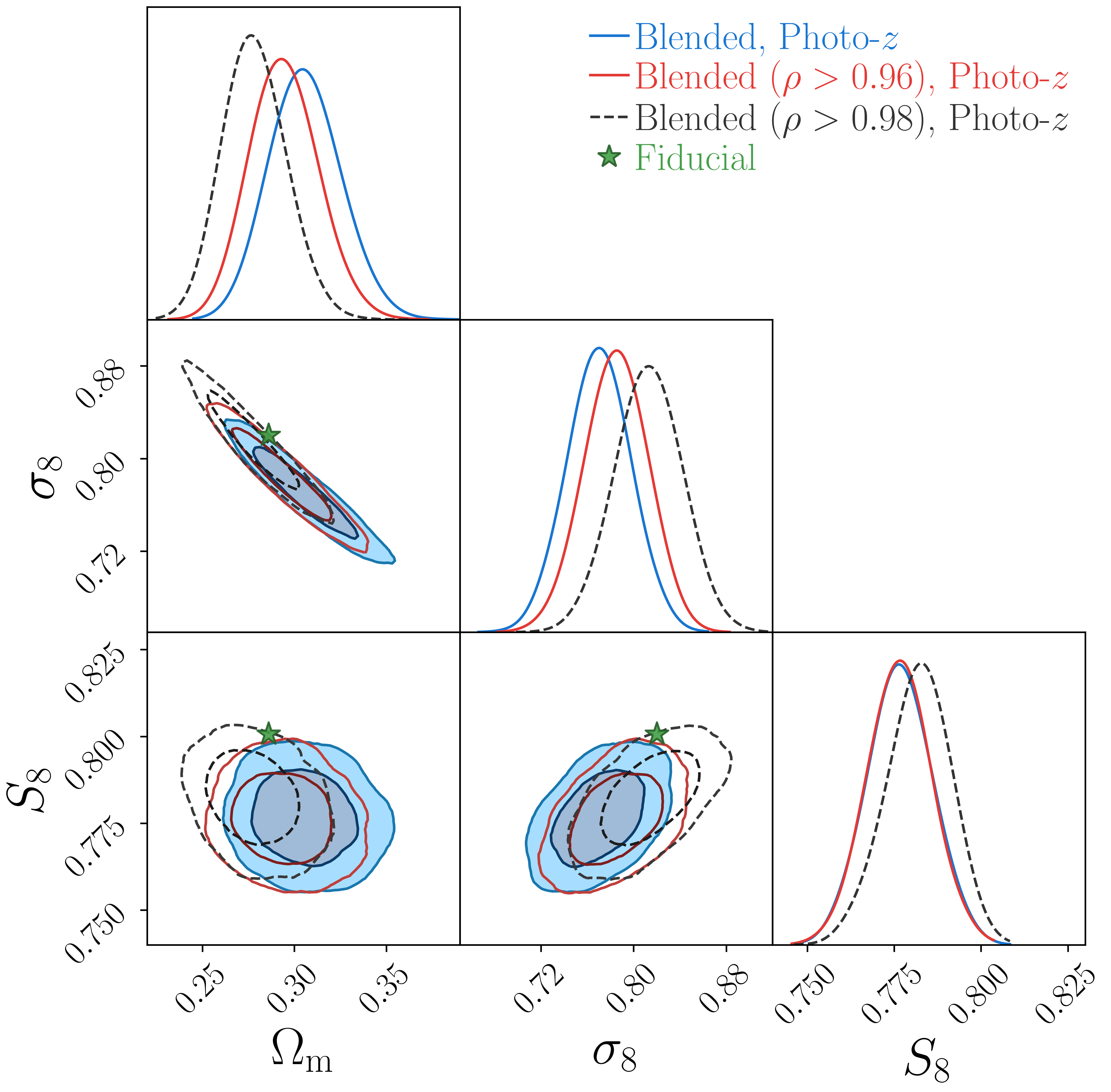}
	\caption{Same as Fig. \ref{cosmoz_rb} but for tomographic bins selected using photometric redshifts. Blending also impacts the photo-$z$'s, resulting in a secondary contribution to the amplitude of the shifts.}
	\label{cosmoz_rb_pz}
\end{figure}

It is therefore useful to ask what the effect on the bias in cosmological inference would be of omitting the catastrophic recognized blends with low purities from the blended source sample. 
The LSST at full depth is expected to have over 30 per cent recognized blends with still a significant dip in surface brightness between the galaxies, as discussed in \cite{scarlet}. We repeat the cosmological analysis that we undertook above for the unrecognized blends, but omit galaxies in the blended scenes by setting a purity cutoff that flags a fraction of the total galaxies in the blended scenes as catastrophic recognized blends. The resulting systematic shift in the error ellipsoid for cosmological parameters is shown in Fig. \ref{cosmoz_rb} for spec-$z$ tomography and in Fig. \ref{cosmoz_rb_pz} for photo-$z$ tomography. 
Comparing this with the shift for unrecognized blends in Figs \ref{cosmoz} and \ref{cosmopz}, we see that the shifts in $S_8$-$\Omega_{\rm m}$ plane are orthogonal. Eliminating recognized blends in two cases where we keep only galaxies with $\rho>0.96$ and $\rho>0.98$ (which removes $\sim 6$ and $\sim 9$ per cent of the galaxies from the cosmology sample, or $\sim 7$ and $\sim 11$ per cent of the galaxies from the source sample, respectively) produces no significant change in $S_8$ for spec-$z$ tomography. However, the extreme case of $\rho>0.98$ results in an $\sim0.006$ increase in $S_8$ for photo-$z$ tomography. 
Interestingly enough, these contour shifts are statistically insignificant when compared with the precision of our analysis. The development of the shifts depends in a non-linear way on the purity threshold for recognized blends. Section \ref{discussion} discusses the advocacy of these results.

The above analysis is repeated using the Gaussian covariance matrix and shows biases to even a lesser extent. As you can see in Figs \ref{cosmoz_rb_gausscov} and \ref{cosmoz_rb_pz_gausscov}, based on the Gaussian covariance matrix as opposed to the jackknife covariance matrix, the results show statistically indistinguishable biases attributed to the omission of low-purity recognized blends.

\bigskip

\begin{table}
\centering
		\begin{tabular}{l r} 
    	    \toprule
    		{\bf Parameter} & {\bf Prior}	\\ \midrule
    		\multicolumn{2}{c}{\it Cosmological} \\ \midrule
    		$\Omega_{\rm m}$ & $\mathcal{U}(0.1, 0.6)$ \\
    		$\sigma_8$ & $\mathcal{U}(0.5, 1.1)$ \\
    		$h$ & $\mathcal{U}(0.5, 1)$\\
    		$\Omega_{\rm b}$ & $\mathcal{U}(0.01, 0.09)$ \\
    		$n_s$ & $\mathcal{U}(0.8, 1.07)$ \\ [2ex]
    		\midrule
    		\multicolumn{2}{c}{\it Systematic} \\ \midrule
    		$m_i$			& $\mathcal{G}(0.02, 0.03)$ \\
    		$\Delta z^1$	& $\mathcal{G}(-0.0064, 0.0003)$ \\
    		$\Delta z^2$	& $\mathcal{G}(-0.0050, 0.0004)$ \\
    		$\Delta z^3$	& $\mathcal{G}(-0.0060, 0.0004)$ \\
    		$\Delta z^4$	& $\mathcal{G}(-0.0038, 0.0005)$ \\ [1ex]
    		\bottomrule
		\end{tabular}
	\caption{The priors used in the analysis for cosmological parameter constraints. We used $\mathcal{U}$ to represent non-informative flat priors in the given range and $\mathcal{G}$ for informative Gaussian priors with the first argument as the mean and the second argument as the dispersion. For the purpose of this study, we fix $w = -1.0$, $\Omega_{\rm k} =  0.0$, $\tau =  0.08$ and consider three massless neutrino species and $N_{\rm eff} = 3.046$ from the input cosmology \citep{derose}. The $m_i$ parameters correspond to the shear calibration and $\Delta z^i$ to the source photo-$z$ bias, with $i = 1...4$ for four tomographic redshift bins. The systematic parameters are set based on the non-blended source sample so that they can serve as our reference to measure potential biases in the blended source sample. Future analyses will likely need to expand these prior ranges further in order to account for the effects of blending.}
	\label{priors}
\end{table}

\begin{figure*}
	\includegraphics[height=180pt]{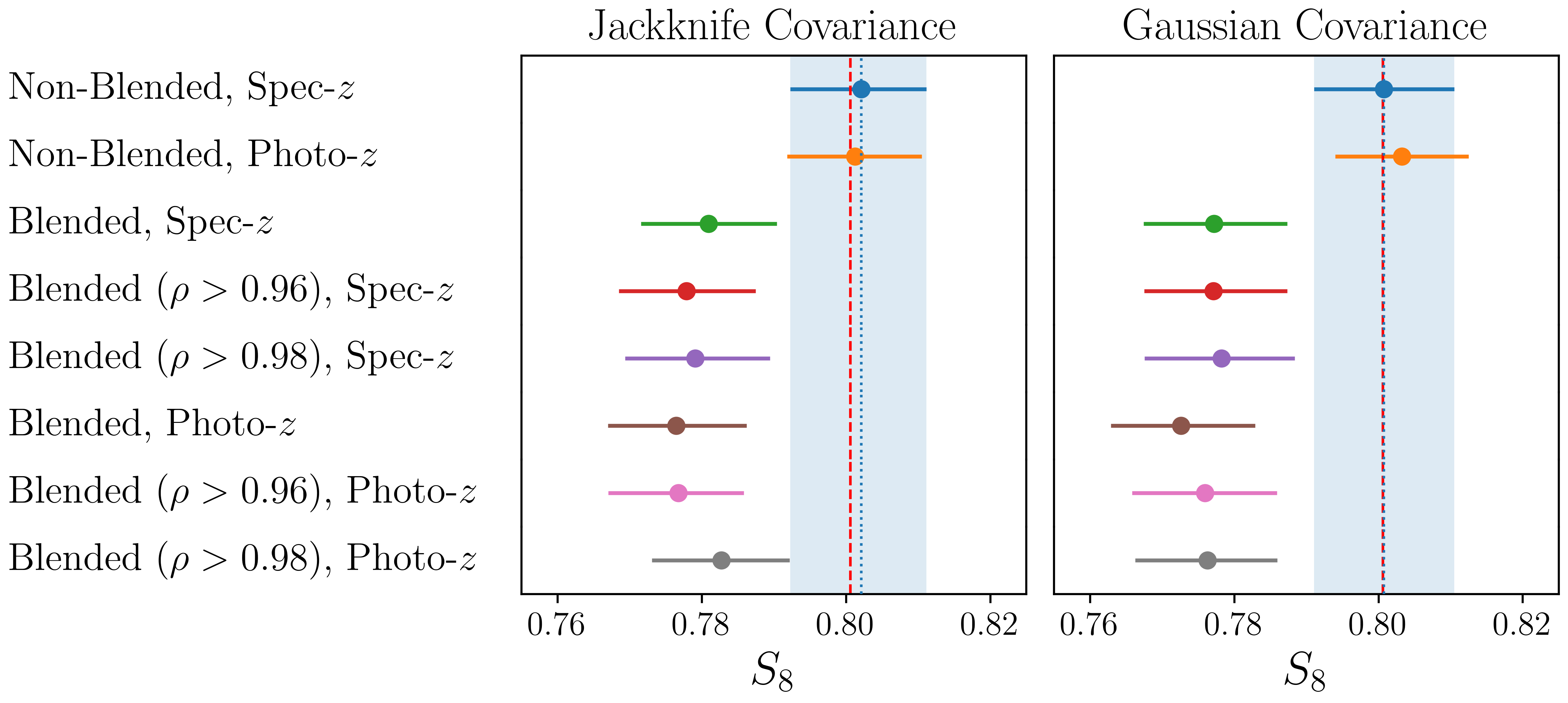}
	\caption{Summary of the $1\sigma$ constraints on the weighted amplitude of matter fluctuations, $S_8=\sigma_8(\Omega_{\rm m}/0.3)^{0.5}$, obtained from different source samples in our cosmic shear analysis. The values on the \textit{left-hand} and \textit{right-hand} panels are estimated by using jackknife and Gaussian covariance matrices, respectively. The shaded $\pm 1\sigma$ intervals in each panel correspond to the non-blended case with spectroscopic tomography containing a dotted line for its mean $S_8$ value. To aid in comparison, the dashed red line indicates the fiducial value taken from the \texttt{Buzzard} simulation. Blending results in a bias of more than the $2\sigma$ statistical error to lower values of $S_8$..}
	\label{S8_summary}
\end{figure*}

\subsection[Summary of the constraints on S8]{Summary of the constraints on $S_8$}\label{s8section}
Fig. \ref{S8_summary} compares the estimated values of the derived parameter $S_8=\sigma_8(\Omega_{\rm m}/0.3)^{0.5}$ and corresponding uncertainties inferred from the jackknife (\textit{left-hand} panel) and Gaussian (\textit{right-hand} panel) covariance matrices. Utilizing two independent types of the covariance matrix, we find that systematic difference in the $S_8$ posteriors due to unrecognized blending is greater than the sum of the associated $1\sigma$ posterior uncertainties. Compared to the Gaussian covariance, the jackknife results are, on the whole, more sensitive to the purity cuts we make to eliminate low-purity blends.

\section{Discussion}\label{discussion}
The driver for the \cite{dawson2016} blend fraction estimate was that LSST will have $\sim0.7~ \rm arcsec$ FWHM PSF, which at the limit of the main 18\,000 square degree survey to its full depth, will result in $\sim 14$ per cent of galaxies in the LSST gold sample ($i<25.3$ where $\rm S/N=20$ for point sources) unrecognizably blended. They consider all galaxies in their space-based ``faint'' sample ($i<27$, which is approximately the LSST limiting magnitude) when determining which of the $i<25.3$ galaxies in their ground-based sample are unrecognized blends\footnote{This means if two galaxies fainter than $i=25.3$ are unrecognizably blended together then the ground-based telescope could detect them as a single galaxy brighter than $i=25.3$ and place that object in the gold sample, which would count towards the unrecognized blend fraction.}.
Naturally, the blend fraction depends on the magnitude limit of the faint sample. For example, the fraction of gold-sample galaxies blended with gold-sample galaxies is far smaller. The science impact can only be assessed via a simulation, which is the subject of this study. The ground-based PSF effect on galaxy apparent shape, noise, and systematic errors due to resulting blends and its science implication is three fold:

\begin{enumerate}[label=(\roman*),leftmargin=*]
\item Distant blended pairs of galaxies have apparent shape that is dominated by the blend (and not the true shape), resulting in noisier shear statistics in a large random sample;
\item Blended galaxies can have incorrectly estimated photo-$z$, generating false correlations across tomographic redshift bins, impacting all cosmology probes including cosmic shear;
\item The ground-based PSF circularizes all galaxy images, yielding systematically biased measured shear for a given shear. This is the shear calibration term \citep{shearm1, shearm2}.
\end{enumerate}
The latter is not the focus of this paper, although it is included and calibrated for in our study and is significant for cosmic shear and galaxy-mass probes. Surveys such as DES \citep{derose2021}, DLS \citep{jz}, HSC \citep{hscshearcal}, and KiDS \citep{heymans2021} have dealt with this via simulations or \textsc{metacalibration} \citep{metacal1, metacal2}. 

It is important that shear errors remain random averaged over all the visits to each sky patch. If not, the systematic error arising from such a situation will limit the cosmological precision achieved. For LSST we need to reduce the residual shear--shear power systematic error to well below the cosmological signal on the angular scales of ten arcminutes to several degrees, which sets a goal of $< 3 \times 10^{-7}$ for the full gold sample, that will be used for cosmological parameter estimation \citep{srd}. 
The main problem arises when one has a mix of effects (i), (ii), and (iii). We need some understanding of the effects (i)+(ii) separately from effect (iii), since (iii) will likely be calibrated to the needed accuracy for real observations.
Effects (i) and (ii) also create a shear multiplicative bias (corrected through the shear calibration parameter) on their own. That is a natural outcome of blending itself, and we need to understand it separately from the PSF rounding. Thus, our simulation is focused on the galaxy--galaxy blending effect itself.

A large survey like LSST will rely on photometric information to estimate redshifts and produce a 3D map of galaxies. To isolate the effects of blending in a catalogue with distances determined by photometric redshifts, we also analyze the enhanced \texttt{Buzzard} catalogue data replacing spectroscopic redshifts with TPZ photometric redshifts.

We estimate that about 12 per cent of the galaxies in our blended catalogue are unrecognized blends that have more than one counterpart in the truth catalogue. We then investigate how systematic errors due to blending (with spec-$z$ or photo-$z$ tomography) propagate to cosmology constraints. In each case (Figs \ref{cosmoz} and \ref{cosmopz} using the jackknife covariances or Figs \ref{cosmoz_gausscov} and \ref{cosmopz_gausscov} using the Gaussian covariances), our simulations show a small reproducible effect on $\Omega_{\rm m}$ and $\sigma_8$ due to blending, and in most cases are consistent with some known mechanism. This suggests that there is some hope of modelling the effects of blending in LSST in order to correct the cosmological parameters to first order, e.g. through an additional shear calibration parameter, $m_{\rm b}$, and perhaps widening of the prior on the shear calibration parameter. This will be aided by cutting the sample in multiple ways in order to validate the models:
a sample with the best seeing; samples with various quality recognized blends removed; a sample absent galaxies having strong colour gradients; and samples with different limiting apparent magnitudes. Finally, the deep drilling fields will enable most of these investigations, cutting ultradeep data in all these ways early in the survey. 

We notice that tightening the priors on the shear calibration parameter (i.e., adopting a smaller $\sigma_{\rm m}$) results in stronger discrepancies between the non-blended and blended cosmology constraints.
On the contrary, relaxing these priors by adopting a larger $\sigma_{\rm m}$ decreases such discrepancies at the expense of lessening the constraining power.

In any case, as summarized in Fig. \ref{S8_summary}, including the effects of blending in our simulation induces a bias in the measured $S_8$ parameter of $\sim 0.025$.  The amplitude of this shift is more than twice the width of the expected 68 per cent confidence interval for measuring $S_8$, and thus a source of uncertainty that needs to be mitigated in future cosmological analyses that utilize deep imaging data. Blending apparently has only a modest effect on photometric redshifts. We believe that this is due to the fact that most of the blends are with at least three times fainter objects that only moderately pollute the flux of the brighter object, and our photo-$z$ algorithm seems to be mostly robust to these flux variations.
 
While methods like \textsc{metacalibration} \citep{metacal1, metacal2} and ``Bayesian Fourier Domain'' (BFD; \citealt{bfd}) are developed to deal with problems in the shape measurement of recognized blends, they do not handle unrecognized blends since, by definition, these blends are not recognized to be two or more galaxies and will not have their shapes separately measured in the first place.
Dealing with the detection bias (e.g., due to unrecognized blends) was the motivation behind developing \textsc{metadetection} \citep{metadet}, a technique that includes object detection in the \textsc{metacalibration} process. It can mitigate the same-redshift unrecognized blending biases in shear measurement as a result of overlapping objects at the same redshift having the same shear. The cross-redshift blending -- which is the most probable scenario -- still remains impossible to handle when the blend is unrecognized.
 
The omission of severe cases of recognized blends (Section \ref{recbl}) is done by applying two lower cuts on the purity of galaxies. As pointed out in Section \ref{blendemu}, all galaxies detected in the blended scenes -- including unrecognized blends -- are assigned purity values and undergo purity cuts.
The cuts are applied to the blended catalogue and introduce slight shifts in the estimated cosmological parameters, more noticeably in the results of photo-$z$ tomography. Overall, the shifts trend towards the fiducial values as we use more stringent purity cuts, except for the estimates of $S_8$ using spec-$z$ tomography with jackknife covariance. By removing the low-purity galaxies with only one counterpart in the non-blended catalogue, we should not observe any visible changes toward the fiducial values in the estimation of our parameters. This is due to the fact that recognized blending bias is not considered when modelling the shape measurement in this study (i.e., purity is not taken into account when adding noise to the shapes), and therefore removing galaxies mentioned above does not remove any blending bias in the shape measurement.
Nevertheless, eliminating low-purity galaxies more than likely leads to the elimination of some ($\sim8$ per cent in the extreme case of eliminating galaxies with $\rho<0.98$) of the unrecognized blends in the blended scenes, thereby reducing the unrecognized blending bias in our cosmology inference. The removal of a fraction of unrecognized blends decreases the rate of outliers with catastrophically wrong photo-$z$ assignments. This may explain why the aforementioned reduction in the unrecognized blending bias (i.e., the overall tendency to approach the fiducial or non-blended values) is more noticeable for the inference with photo-$z$ tomography.

The future lies in increasing both the depth and area of sky surveys. The number of galaxies that we used for this simulation ($\sim$ 0.5 billion as sources) is still smaller than the number expected in the LSST gold sample ($\sim$ 3 billion), and we expect an increase in statistical accuracy by using a larger sample. This also enables the model validation tests via multiple cuts on the catalogue.
Even so, there are sufficient galaxies present in our sample such that the systematic effects we seek are more significant than the statistical noise due to the limited sample size.   We also have no artefacts such as bright stars, varying sky background, dust, or changing and non-circular PSF. Finally, we include no camera effects that could print through to the co-add and the catalogue. We are focused on blending -- our goal is to isolate the effects of galaxy blends.

We analysed a sample equivalent of 1.3 yr of LSST operation. The ultimate 10-yr depth of LSST exacerbates the blending impact, for which we will need additional techniques to model it out.  This study yields an estimate of the effect of galaxy blending and prospects for improvement. Due to the shared 3D large-scale dark matter structure, there is hope that a joint analysis of multiple independent probes will suppress the 2D blending systematic errors in cosmology. This so-called $3\times2 {\rm pt}$ analysis -- which includes galaxy clustering and galaxy-galaxy lensing (a.k.a. galaxy-mass correlation) in addition to the cosmic shear studied in this paper -- will be carried out in a follow-on study.

\section*{Acknowledgments}
This paper has undergone internal review in the LSST Dark Energy Science Collaboration. We thank the internal reviewers, Bob Armstrong, Patricia Burchat, and Javier S\'{a}nchez for their helpful comments.

We acknowledge support from NSF/AURA/LSST grant N56981CC and DOE grant DE-SC0009999. 
This work used the Extreme Science and Engineering Discovery Environment (XSEDE; \citealt{xsede}) Comet and Expanse clusters at the San Diego Supercomputer Center (SDSC) through XSEDE Research Allocation (under award number AST200036) which is supported by National Science Foundation grant number ACI-1548562.
This work also benefited from the HPC at UC program project DDP340 through the SDSC Comet cluster. This research used resources of the National Energy Research Scientific Computing Center (NERSC), a U.S. Department of Energy Office of Science User Facility located at Lawrence Berkeley National Laboratory, operated under Contract No. DE-AC02-05CH11231.
We gratefully acknowledge support from the CNRS/IN2P3 Computing Center (Lyon, France) for providing computing and data-processing resources needed for this work.

We thank Alexie Leauthaud for providing the weak lensing catalogue for COSMOS which was used for testing the \texttt{BlendSim} pipeline. COSMOS data is based on data products from observations made with ESO Telescopes at the La Silla Paranal Observatory under ESO programme ID 179.A-2005 and on data products produced by TERAPIX and the Cambridge Astronomy Survey Unit on behalf of the UltraVISTA consortium.
EN would like to thank Mike Jarvis for his advice on weak lensing analysis and using \texttt{TreeCorr} and \texttt{GalSim}, Joe DeRose for his advice on using the \texttt{Buzzard} simulation, Patricia Burchat and Sowmya Kamath on blending emulation, Sukhdeep Singh on the Gaussian covariance matrices, Yao-Yuan Mao on using the \texttt{GCRCatalogs} package, and Javier S\'{a}nchez, James Jee, Imran Hasan, Joe Zuntz, Peter Melchior, David Kirkby, Andrew Bradshaw, Mohsen Khorasani, Alan Heavens, David Alonso, Ali Mahzarnia, Josh Meyers, Erin Sheldon, Matias Carrasco Kind, Will Dawson, Michael Schneider, Andrew Hearin, and Stephanie Escoffier for helpful discussions. EN also thanks Mahidhar Tatineni for his help with the Comet and Expanse clusters at SDSC.

The DESC acknowledges ongoing support from the Institut National de 
Physique Nucl\'eaire et de Physique des Particules in France; the 
Science \& Technology Facilities Council in the United Kingdom; and the
Department of Energy, the National Science Foundation, and the LSST 
Corporation in the United States.  DESC uses resources of the IN2P3 
Computing Center (CC-IN2P3--Lyon/Villeurbanne, France) funded by the 
Centre National de la Recherche Scientifique; the National Energy 
Research Scientific Computing Center, a DOE Office of Science User 
Facility supported by the Office of Science of the U.S.\ Department of
Energy under Contract No.\ DE-AC02-05CH11231; STFC DiRAC HPC Facilities, 
funded by UK BIS National E-infrastructure capital grants; and the UK 
particle physics grid, supported by the GridPP Collaboration.  This 
work was performed in part under DOE Contract DE-AC02-76SF00515.

We made use of the following software packages in addition to those already cited in the text: \texttt{Numpy} and \texttt{Scipy} \citep{numpyscipy}, \texttt{Astropy} \citep{astropy2013, astropy2018}, \texttt{APLpy} \citep{aplpy}, \texttt{Matplotlib} \citep{matplotlib}, \texttt{Pandas}  \citep{pandas}, \texttt{Healpy} \citep{healpix19}, \texttt{MPI4Py} \citep{mpi4py05}, and \texttt{NetworkX} \citep{networkx}.

\section*{Data Availability}
The pipeline written for this project is available at \url{https://github.com/enourbakhsh/BlendSim}. The processed data underlying this paper will be available upon request. This article uses data provided by the \texttt{Buzzard Flock} simulation group \citep{derose}, which will also be available upon request, and with the permission of the simulation group.



\appendix
\addcontentsline{toc}{section}{Appendices}

\section{The effect of non-Gaussian likelihood}\label{ngl}
The inverse of the covariance matrix is used in the likelihood that infers cosmology from the data. Due to the large dimensionality of the covariance matrix in current cosmological studies and the fact that we have a limited number of pseudo-samples to internally estimate the covariance matrix, we often end up with a noisy covariance matrix e.g. from jackknife resampling. These noisy covariance matrices lead to biased estimates of the inverse covariance thereby biasing our estimates for cosmological parameters.

Efforts have been made to correct for this effect due to a finite number of samples from simulations of mock data. \cite{hartlap07} tries to de-bias the inverse of the covariance matrix $C^{-1}$ while keeping the likelihood Gaussian. The corrected inverse is estimated by
\begin{equation}\label{hc}
    C^{-1}_{\rm corrected} = \frac{N_{\rm sim}-N_{\rm bin}-2}{N_{\rm sim}-1} ~ C^{-1}
\end{equation}
where $N_{\rm bin}$ is the number of angular bins and $N_{\rm sim}$ is the number of samples the covariance is generated from. However, \cite{sh16} noted that the likelihood is more consistent with a $t$-distribution rather than a Gaussian due to marginalization over the noisy covariance matrix.

\begin{figure*}
	\centering
    \includegraphics[width=\textwidth]{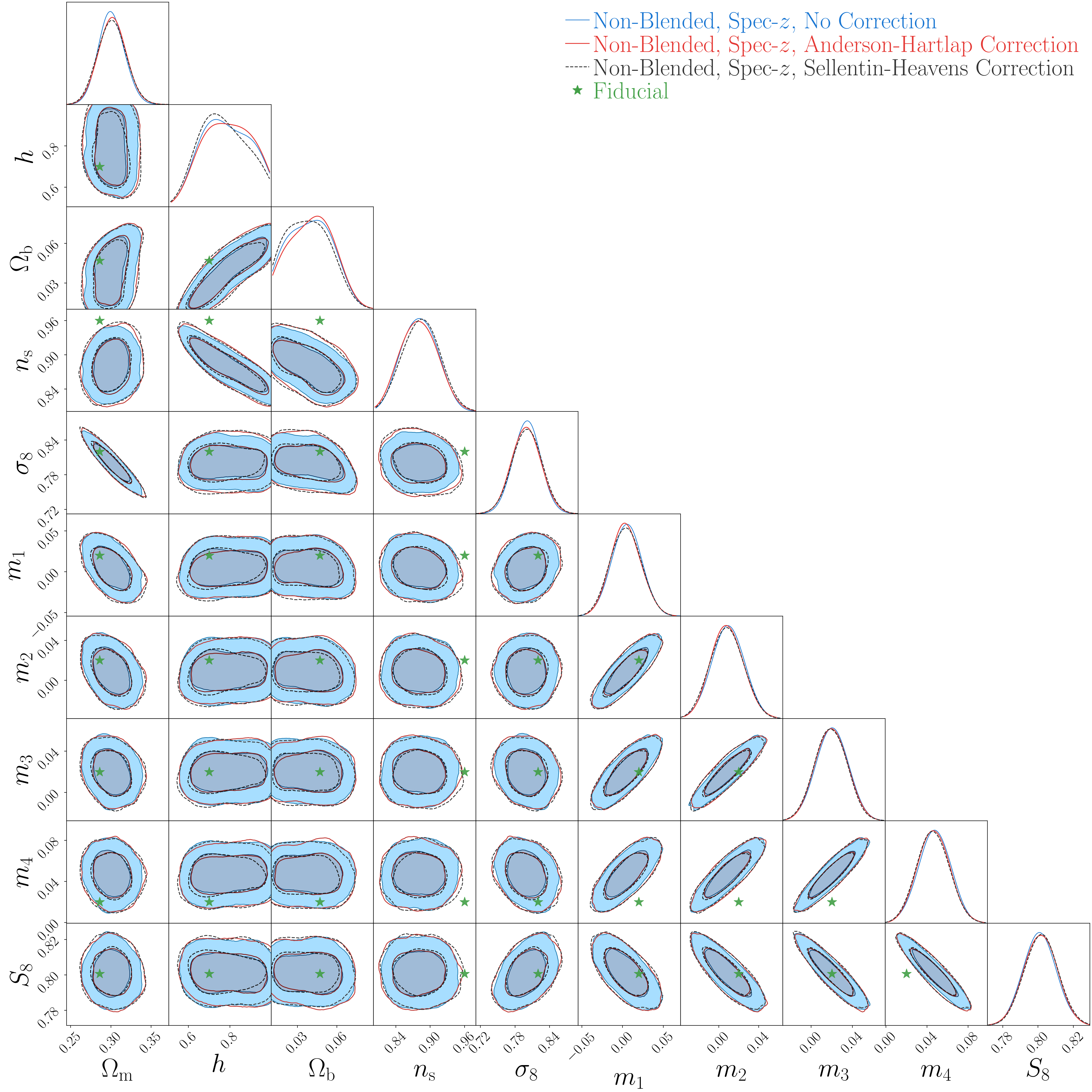}
	\caption{Cosmological parameter estimation based on cosmic shear in the non-blended spectroscopic source sample and two different corrections, Anderson--Hartlap and Sellentin--Heavens, on the covariance matrix produced by the jackknife estimator. The contours of the chains with not-corrected, Anderson--Hartlap corrected and Sellentin--Heavens corrected covariance matrices are displayed in filled blue, unfilled red (solid line), and unfilled black (dashed line), respectively. In each case, the contours drawn correspond to $1\sigma$ (68 per cent) and $2\sigma$ (95 per cent) confidence levels. The green star marks the fiducial parameters. The three sets of posteriors do not differ significantly except that the non-corrected case imposes tighter constraints for $\Omega_{\rm m}$ and $\sigma_8$.}
	\label{covarcomp} 
\end{figure*}

In order to investigate this effect in our data, we have calculated the correction both ways using \texttt{CosmoSIS}. Fig. \ref{covarcomp} demonstrates that the differences between these two corrections (and compared to the case without a correction) are insignificant for our data and in fact, are mostly within the numerical noise. This is due to the fact that $N_{\rm sim}$ is sufficiently large in our sample. Note that the cosmic shear probe does not place strong constraints on $h$, $\Omega_{\rm b}$ and $n_{\rm s}$, as seen from this figure; however, they will become more relevant in a full $3\times2{\rm pt}$ analysis. The fiducial points (green stars) for the shear calibration parameters are not supplied by the \texttt{Buzzard} simulation, contrary to the cosmological parameters that are used to generate the simulation. They all display a fixed ``non-tomographic'' Gaussian prior mean of 0.02 which was the basis for the sampling. We, therefore, do not expect them to remain constant across different tomographic redshift bins.

We choose to use the Sellentin--Heavens correction throughout this paper with a caveat that the effective number of independent samples, which is assumed to be $N_{\rm sim}=2,000$, may not be known for jackknife covariances. However, it might be better to use a correction on a jackknife covariance than to not use any correction at all. The $t$-distribution will widen the likelihood tails, and that certainly has to be better (especially if one is investigating few-sigma tensions) than the Gaussian, which cuts off sharply. For that reason, the Hartlap correction (equation \ref{hc}) would not be a viable idea.

\section{Results with the Gaussian covariance matrix}\label{cosmosis_gcov}
\begin{figure}
	\centering
	\includegraphics[width=\linewidth]{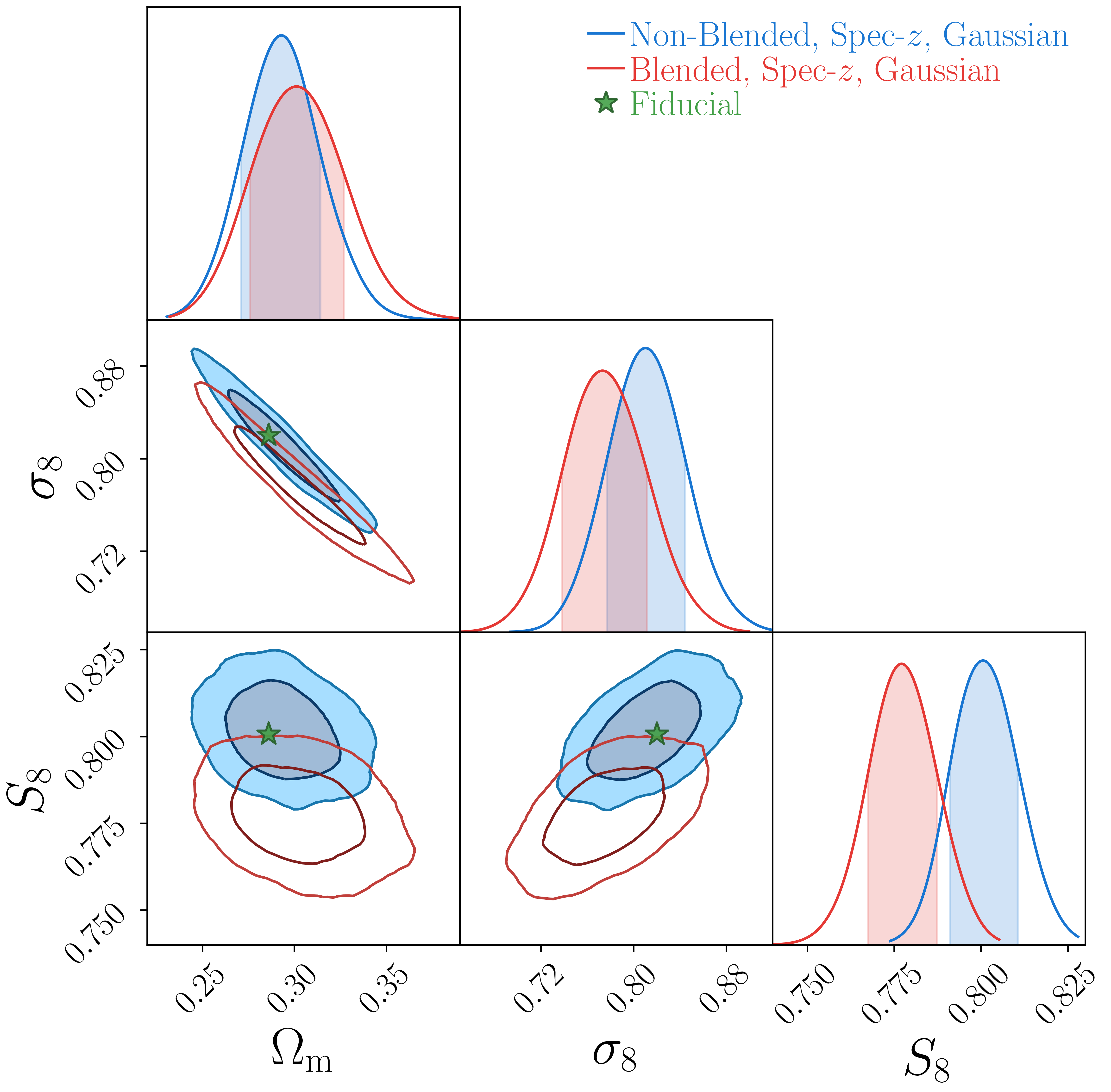}
	\caption{Same as Fig. \ref{cosmoz} but with a Gaussian covariance matrix for each of the two cases.}
	\label{cosmoz_gausscov}
\end{figure}

\begin{figure}
	\centering
	\includegraphics[width=\linewidth]{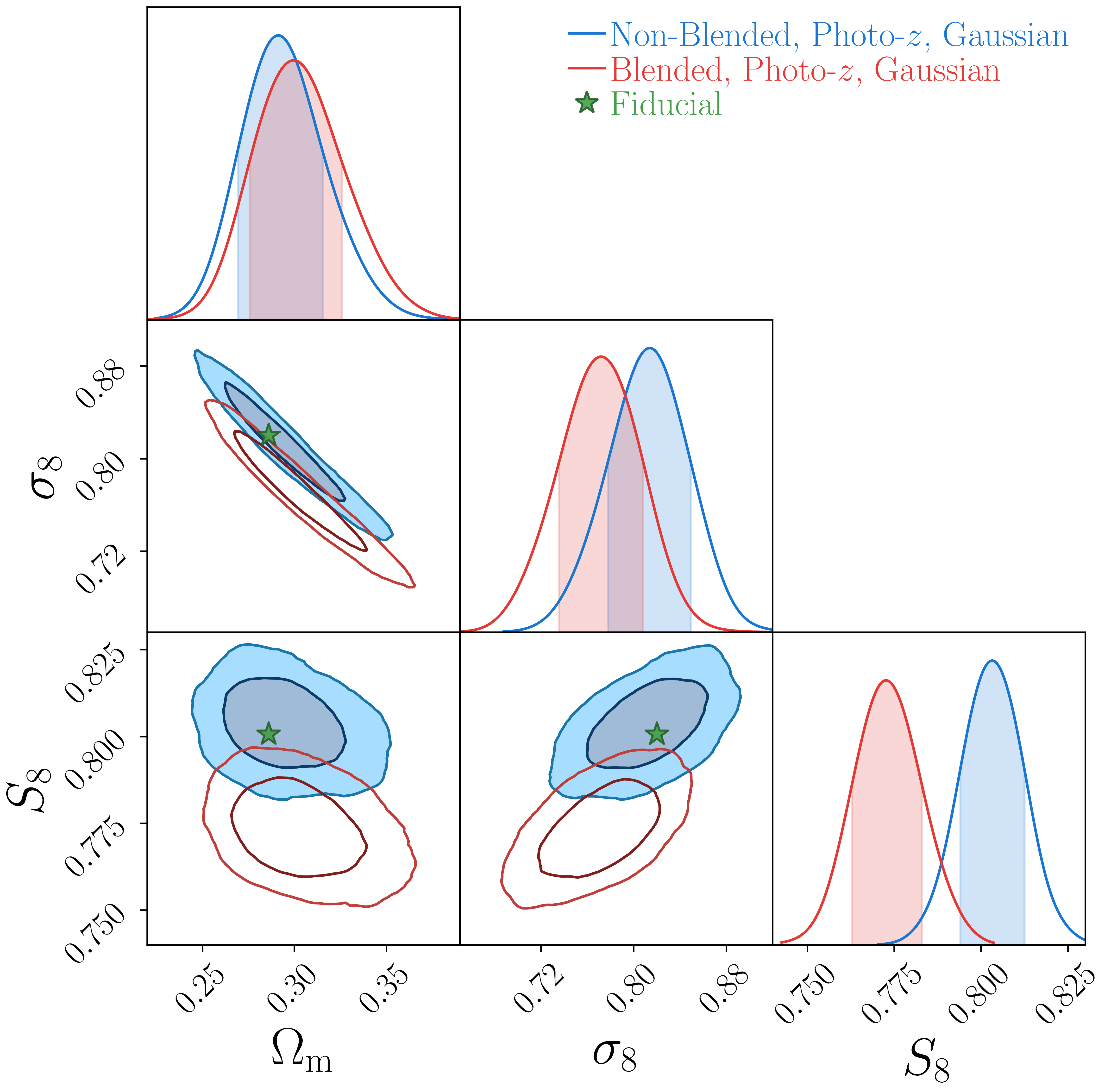}
	\caption{Same as Fig. \ref{cosmopz} but with a Gaussian covariance matrix for each of the two cases.}
	\label{cosmopz_gausscov}
\end{figure}

\begin{figure}
	\centering
    \includegraphics[width=\linewidth]{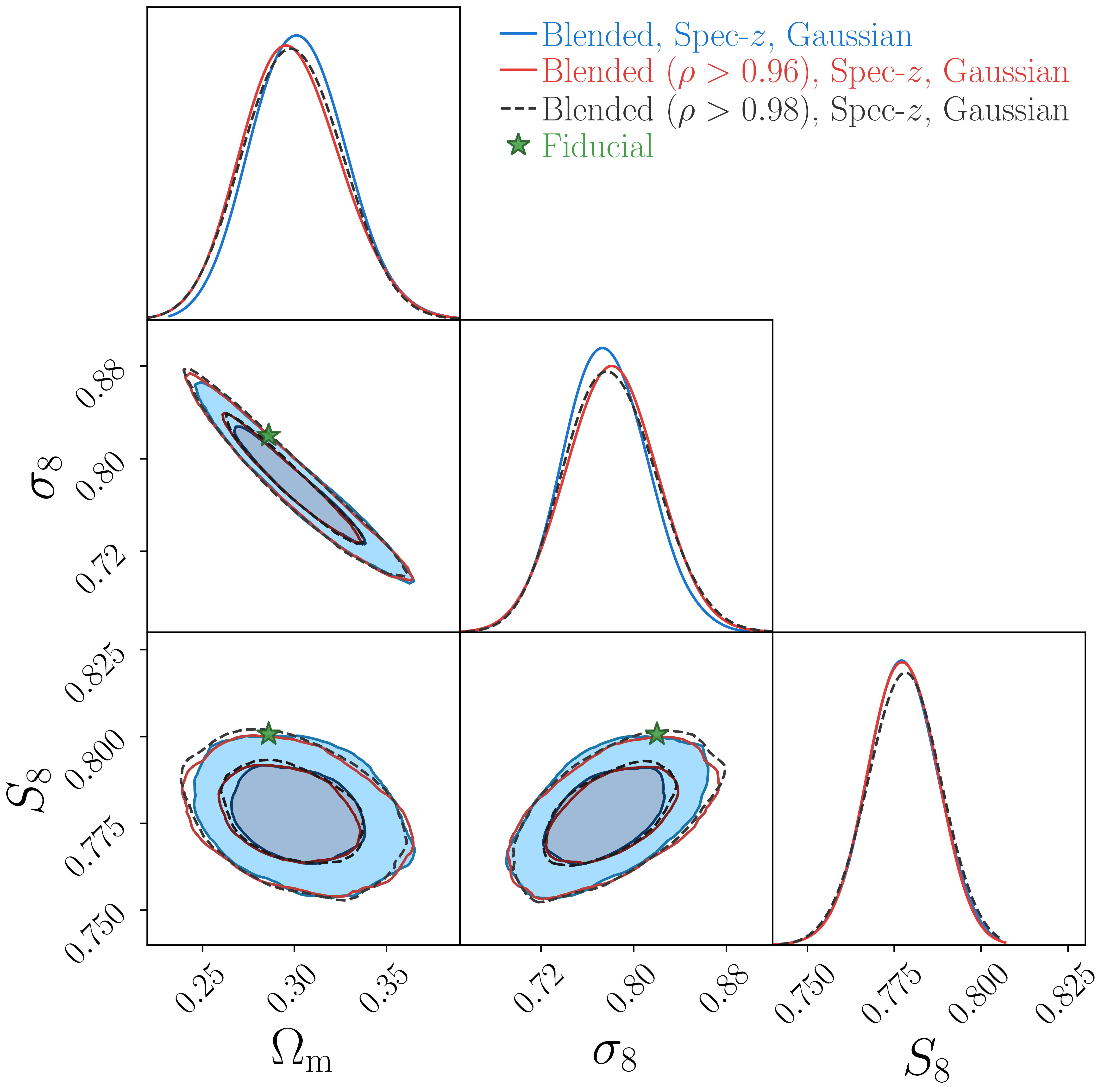}
	\caption{Same as Fig. \ref{cosmoz_rb} but with a Gaussian covariance matrix for each of the three cases.}
	\label{cosmoz_rb_gausscov} 
\end{figure}

\begin{figure}
	\centering
    \includegraphics[width=\linewidth]{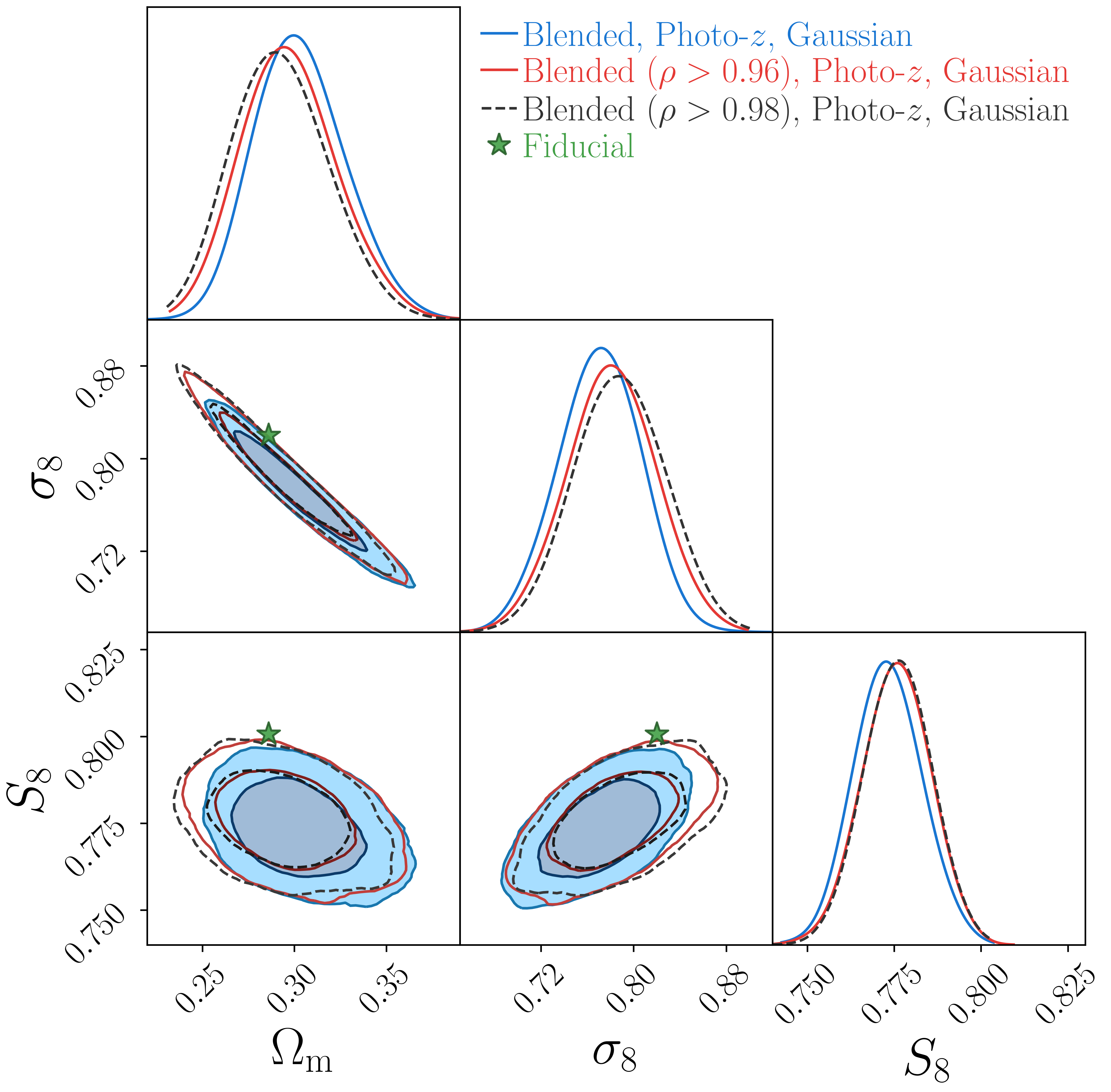}
	\caption{Same as Fig. \ref{cosmoz_rb_pz} but with a Gaussian covariance matrix for each of the three cases.}
	\label{cosmoz_rb_pz_gausscov} 
\end{figure}

We repeat our analysis with the analytical Gaussian covariance matrix described in Section \ref{covariance}, instead of the jackknife covariance matrix. The results are shown in Figs \ref{cosmoz_gausscov}--\ref{cosmoz_rb_pz_gausscov}.

\label{lastpage}
\end{document}